\documentclass[12pt]{article}
\usepackage{amssymb}
\usepackage{amsmath}
\usepackage{bbm}
\usepackage[space]{cite}
\usepackage{graphicx}
\usepackage{mwe}

\numberwithin{equation}{section}

\newcommand{\be}{\begin{equation}}
\newcommand{\ee}{\end{equation}}
\newcommand{\non}{\nonumber}
\newcommand{\id}{\mathbb{I}}
\newcommand{\B}{\mathbb{B}}
\newcommand{\C}{\mathbb{C}}
\newcommand{\E}{\mathbb{E}}
\newcommand{\Hb}{\mathbb{H}}
\newcommand{\K}{\mathbb{K}}
\newcommand{\M}{\mathbb{M}}
\newcommand{\R}{\mathbb{R}}
\newcommand{\T}{\mathbb{T}}
\newcommand{\tb}{\mathbbm{t}}
\newcommand{\eb}{\mathbf{e}}
\newcommand{\tr}{\mathop{\rm tr}\nolimits}
\newcommand{\diag}{\mathop{\rm diag}\nolimits}

\usepackage[usenames,dvipsnames]{color}

\begin{document}

\begin{titlepage}
\strut\hfill UMTG--308
\vspace{.5in}
\begin{center}

{\LARGE  Factorization identities and algebraic Bethe ansatz\\[0.2cm]
for $D^{(2)}_{2}$ models}\\
\vspace{1in}
\large 
Rafael I. Nepomechie \footnote{Physics Department,
            P.O. Box 248046, University of Miami, Coral Gables, FL 33124 USA, nepomechie@miami.edu}
and Ana L. Retore  \footnote{School of Mathematics \& Hamilton 
Mathematics Institute, Trinity College Dublin, Dublin, Ireland, retorea@maths.tcd.ie}

\end{center}

\vspace{.5in}

\begin{abstract}
We express $D^{(2)}_{2}$ transfer matrices as products of
$A^{(1)}_{1}$ transfer matrices, for both closed and open spin chains.
We use these relations, which we call factorization identities, to
solve the models by algebraic Bethe ansatz.  We also formulate and
solve a new integrable XXZ-like open spin chain with an even number of
sites that depends on a continuous parameter, which we interpret as
the rapidity of the boundary.
\end{abstract}

\end{titlepage}

\setcounter{footnote}{0}

\section{Introduction}\label{sec:intro}

The antiferromagnetic Potts model and the staggered six-vertex model
\cite{Potts:1952, Temperley:1971iq, Baxter:1976, Baxter:1982b,
Saleur:1990vd, Jacobsen:2005xz, Ikhlef:2008zz, Ikhlef:2011ay,
Candu:2013fva, Frahm:2013cma, Bazhanov:2019xvy} have recently been
shown \cite{Robertson:2020eri} to be related to the $D^{(2)}_{2}$
R-matrix \cite{Bazhanov:1984gu, Jimbo:1985ua, Bazhanov:1986mu}.  Even
more recently, an open $D^{(2)}_{2}$ spin chain with a particular
integrable boundary condition has been shown \cite{Robertson:2020imc} to
have as its continuum limit a \emph{non-compact} boundary conformal
field theory, which possesses a continuous spectrum of conformal 
dimensions; it is closely related to the $SL(2,\mathbb{R})/U(1)$
Euclidean black hole \cite{Witten:1991yr, Dijkgraaf:1991ba, Maldacena:2000hw,
Maldacena:2000kv,Hanany:2002ev}, see also \cite{Bazhanov:2020dlm, Bazhanov:2020fbp,
Bazhanov:2020uju}.

Here we express the $D^{(2)}_{2}$ transfer matrices for the open spin 
chains considered in \cite{Robertson:2020eri} and \cite{Robertson:2020imc} as
products of $A^{(1)}_{1}$ transfer matrices.  We then use these
relations, which we call factorization identities, to solve the models
by algebraic Bethe ansatz.  In particular, we construct the models'
Bethe states, which had not been known, that would be needed to
compute scalar products and correlation functions.  Moreover, we
prove previously-proposed expressions for the models' eigenvalues and
Bethe equations \cite{Robertson:2020imc, Reshetikhin:1987, 
Martins:2000xie, Nepomechie:2017hgw, Nepomechie:2018nvl}.  
The interesting degeneracies exhibited by these
models are also explained.

In the course of this work, we also formulate and solve a new
integrable XXZ-like open spin chain, which depends on a continuous
parameter. 
We interpret this parameter as the rapidity of the
boundary.  We conjecture that this model, like the one in 
\cite{Robertson:2020imc}, has a non-compact continuum limit.

This paper is structured as follows.  In Sec.  \ref{sec:Rmat}, we give
an exact formulation (\ref{D22Ra})-(\ref{D22Rb}) of the
factorization \cite{Robertson:2020eri} of the $D^{(2)}_{2}$ R-matrix
in terms of $A^{(1)}_{1}$ R-matrices.  Sec.  \ref{sec:closed} is
devoted to the closed $D^{(2)}_{2}$ spin chain.  We use the factorization
of the R-matrix to derive the factorization identity
(\ref{tttrltna})-(\ref{tttrltnb}), which expresses the $D^{(2)}_{2}$
transfer matrix as a product of $A^{(1)}_{1}$ transfer matrices.  We
then use this identity to solve the model by means of algebraic Bethe
ansatz. Since these computations are straightforward, they
may serve as a warm-up exercise for the parallel -- but 
technically more complicated -- computations that follow. 

The heart of this paper is Sec.  \ref{sec:open}, where we consider
open $D^{(2)}_{2}$ chains with two different sets of integrable
boundary conditions, corresponding to the two possible values (namely,
0 and 1) of a certain parameter $\varepsilon$.  We consider first the
case $\varepsilon=1$, which was studied in
\cite{Robertson:2020imc}.  The factorization identity
(\ref{tttrltneps1a})-(\ref{tttrltneps1b}), whose derivation is
presented in Appendix \ref{sec:factopen}, involves a novel
$A^{(1)}_{1}$ transfer matrix (\ref{teps1}).  It is a special case of
the more general transfer matrix (\ref{topen}), which depends on an arbitrary parameter
$u_{0}$ that (as remarked above) we interpret 
as the rapidity of the boundary. We solve 
the general model by algebraic Bethe ansatz, from which we then 
extract the solution for the case $\varepsilon=1$. We treat the case 
$\varepsilon=0$, which was studied in \cite{Robertson:2020eri}, in a similar way.
Its factorization identity (\ref{tttrltneps0a})-(\ref{tttrltneps0b}), whose derivation is
also presented in Appendix \ref{sec:factopen}, 
involves a conventional $A^{(1)}_{1}$ transfer matrix (\ref{teps0}), 
corresponding to $u_{0}=0$. 
In Sec. \ref{sec:gen}, we point out a special case of the model  
(\ref{topen}) with a local Hamiltonian 
for general values of $u_{0}$.
We conclude with a brief discussion of 
our results in Sec. \ref{sec:discuss}.

\section{Product-form R-matrices}\label{sec:Rmat}

We begin this section by reviewing in Sec.  \ref{sec:general} a well-known general
recipe for constructing an R-matrix by forming suitable tensor products of
multiple copies of a more elementary R-matrix.  We actually need
a (perhaps less familiar) generalization of this construction, namely
(\ref{Rprodgen}). Indeed, in Sec. \ref{sec:D22}, we see that the recent 
factorization \cite{Robertson:2020eri} of the $D^{(2)}_{2}$ R-matrix in 
terms of $A^{(1)}_{1}$ R-matrices is precisely of this type, up to a 
similarity transformation. The result (\ref{D22Ra})-(\ref{D22Rb}) is 
the basis for all the factorization identities that we will derive in this 
paper, which express $D^{(2)}_{2}$ transfer matrices as products of 
$A^{(1)}_{1}$ transfer matrices.

\subsection{Generalities}\label{sec:general}

\begin{figure}[htb]
    \centering
    \begin{minipage}{0.40\textwidth}
        \centering
        \includegraphics[width=0.4\textwidth]{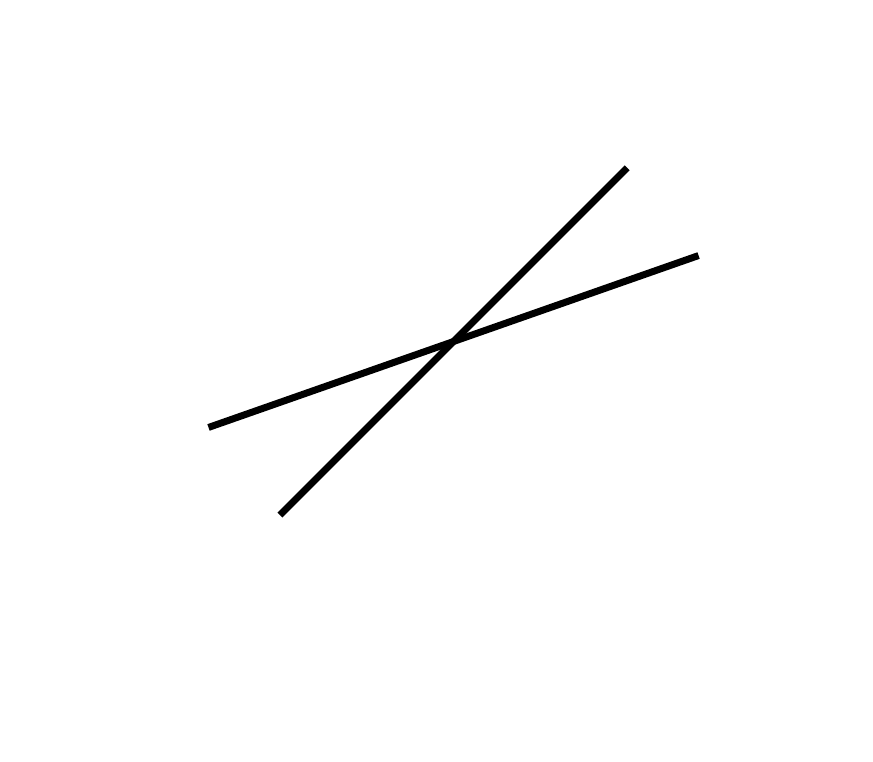} 
        \caption{$R(u)$}
		\label{fig:R}
    \end{minipage}\hfill
    \begin{minipage}{0.40\textwidth}
        \centering
        \includegraphics[width=0.8\textwidth]{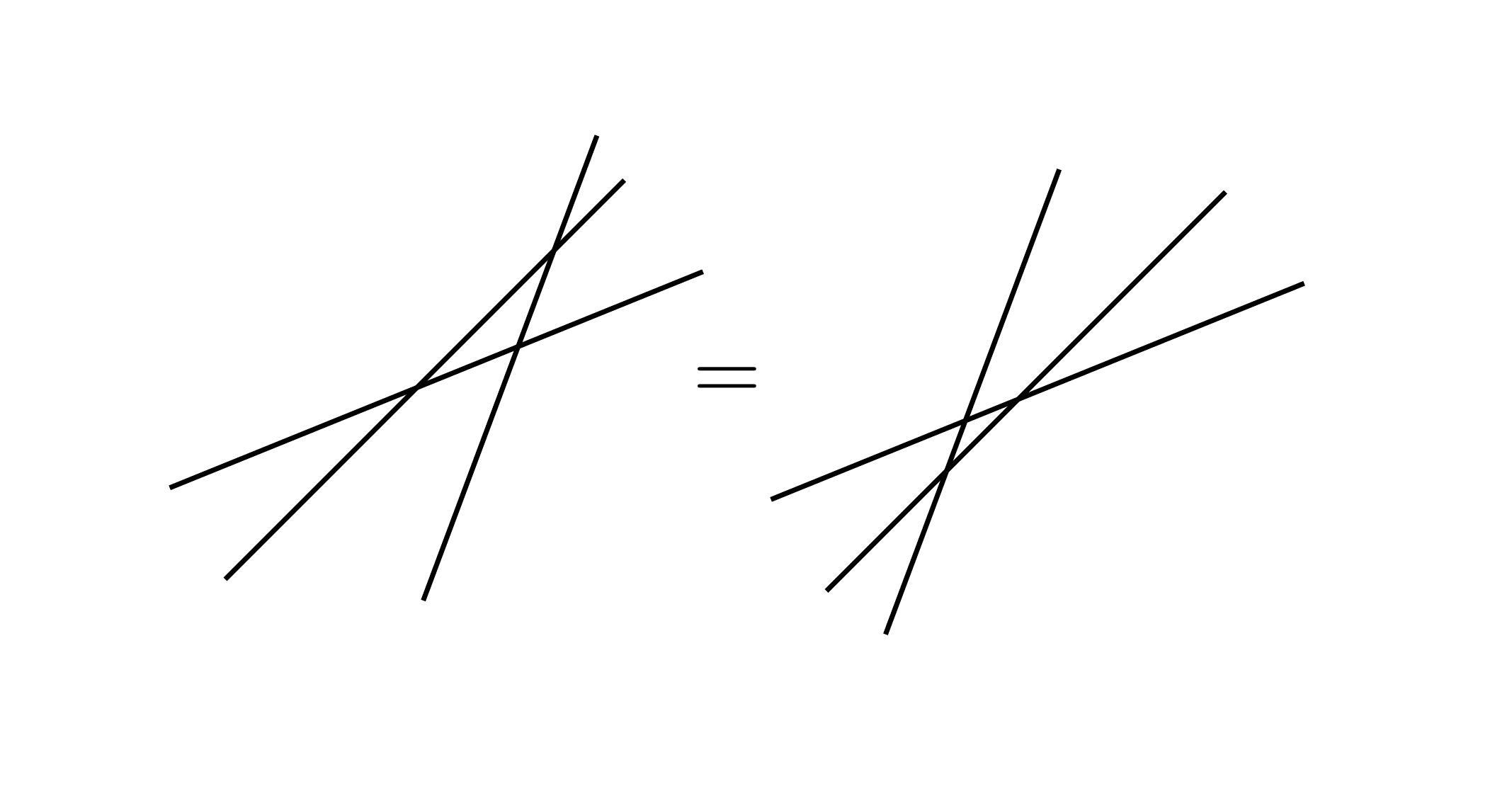}
        \caption{YBE for $R(u)$}
		\label{fig:YBE}
    \end{minipage}
\end{figure}

Consider a solution $R(u)$ of the Yang-Baxter equation (YBE)
\be
R_{12}(u - v)\,  R_{13}(u)\, R_{23}(v) = R_{23}(v)\,  R_{13}(u)\, 
R_{12}(u - v) \,.
\label{YBE}
\ee
As usual, $R(u)$ is a $d^{2} \times d^{2}$ matrix that maps ${\cal V} \otimes {\cal V} \mapsto {\cal V} 
\otimes {\cal V}$, where ${\cal V}$ is a $d$-dimensional vector 
space. In (\ref{YBE}), $R_{12} = R \otimes \id\,, R_{23} = \id \otimes R\,, R_{13} = 
{\cal P}_{23}  R_{12} {\cal P}_{23}$, where here $\id$ is the identity 
matrix on ${\cal V}$ (below, by abuse of notation, $\id$ may denote the identity matrix 
on more than one copy of ${\cal V}$, depending on the context), 
and ${\cal P}$ is the permutation matrix on ${\cal V} \otimes {\cal V}$
\be
{\cal P}=\sum_{a, b = 1}^d e_{a b}\otimes e_{b a} \,,
\ee
where $e_{a b}$ are the $d \times d$ elementary 
matrices with elements $(e_{a b})_{ij} = 
\delta_{a, i} \delta_{b, j}$. As is well known, the R-matrix can be 
usefully represented graphically by one pair of lines that cross, as 
shown in Fig. \ref{fig:R}; hence 
the YBE (\ref{YBE}) is represented using three lines, as shown in 
Fig. \ref{fig:YBE}. 

\begin{figure}[htb]
	\centering
	\includegraphics[width=0.3\textwidth]{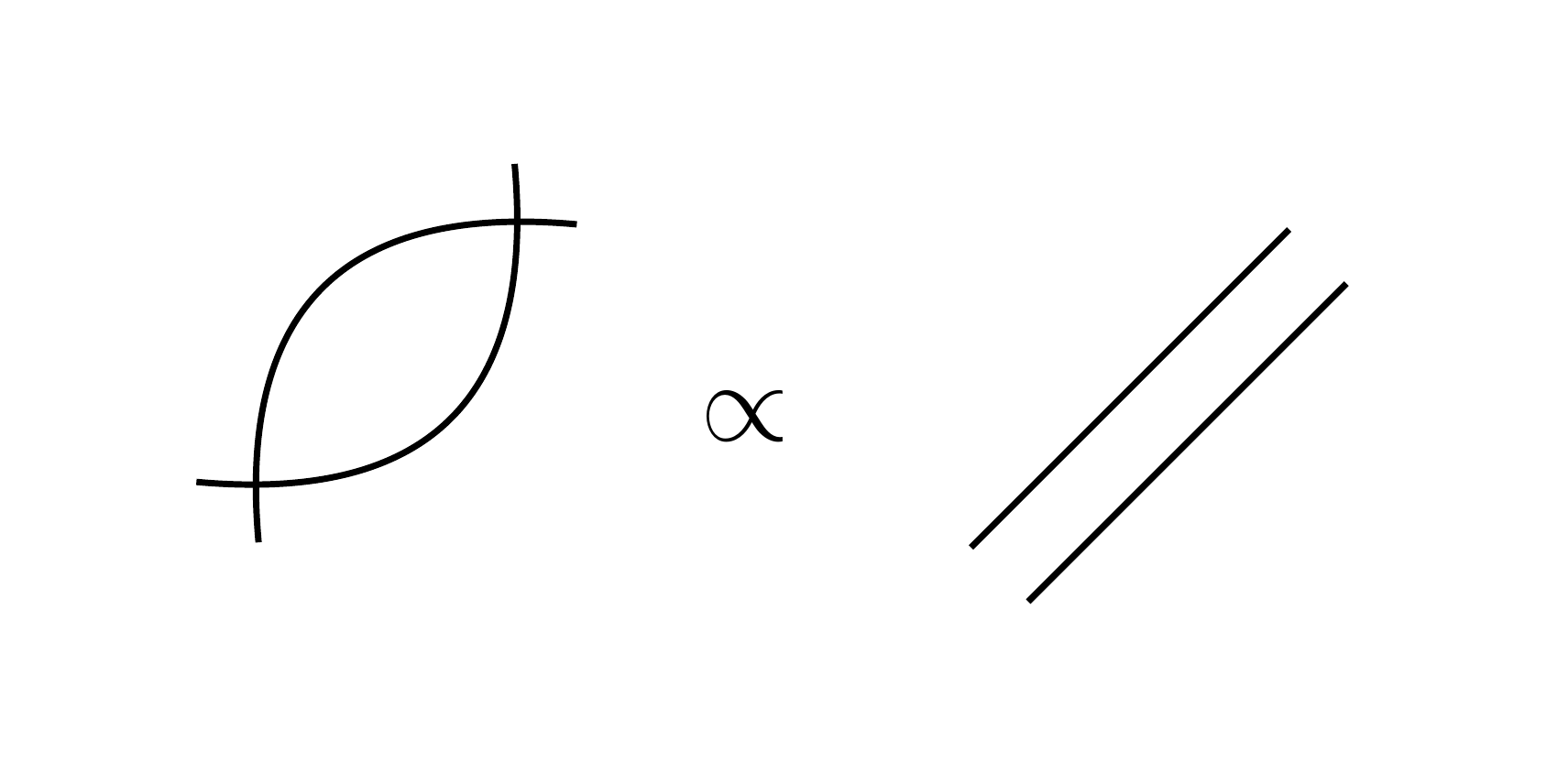}
	\caption{Unitarity}
	\label{fig:unitarity}
\end{figure}

\noindent
We assume that the R-matrix is regular 
\be
R(0) \propto {\cal P} \,,
\label{reg}
\ee
and unitary
\be
R_{12}(u)\, R_{21}(-u) \propto \id \,,
\label{unitarity}
\ee
where $R_{21} = {\cal P}_{12}\, R_{12}\, {\cal P}_{12}$. We use the 
symbol $\propto$ to denote equality up to a scalar factor. The latter 
can be represented graphically as in Fig. \ref{fig:unitarity}.

\begin{figure}[htb]
    \centering
    \begin{minipage}{0.40\textwidth}
        \centering
        \includegraphics[width=0.4\textwidth]{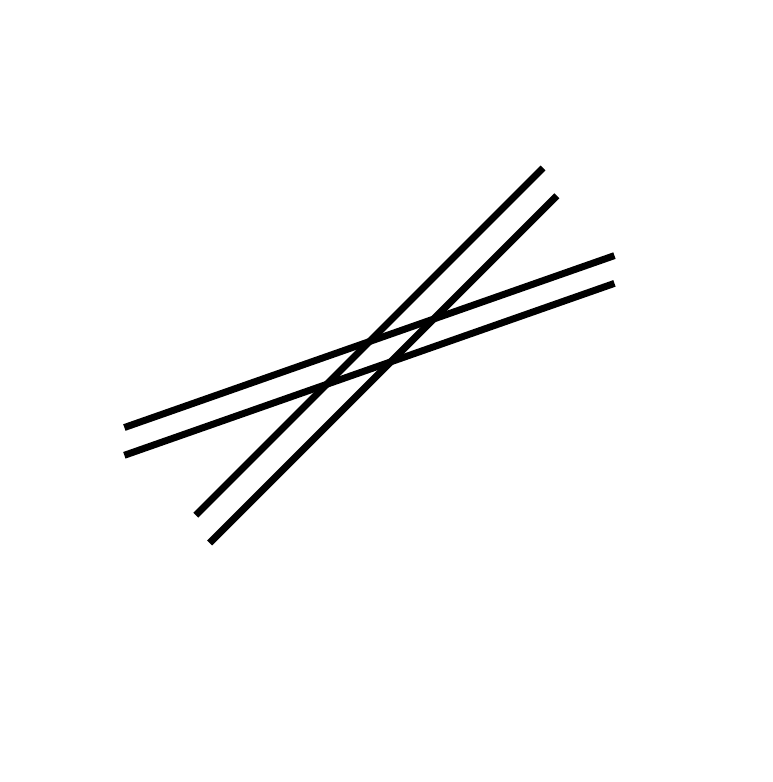} 
        \caption{$\R(u)$ in Eq. (\ref{Rprod})}
		\label{fig:RR}
    \end{minipage}\hfill
    \begin{minipage}{0.40\textwidth}
        \centering
        \includegraphics[width=0.8\textwidth]{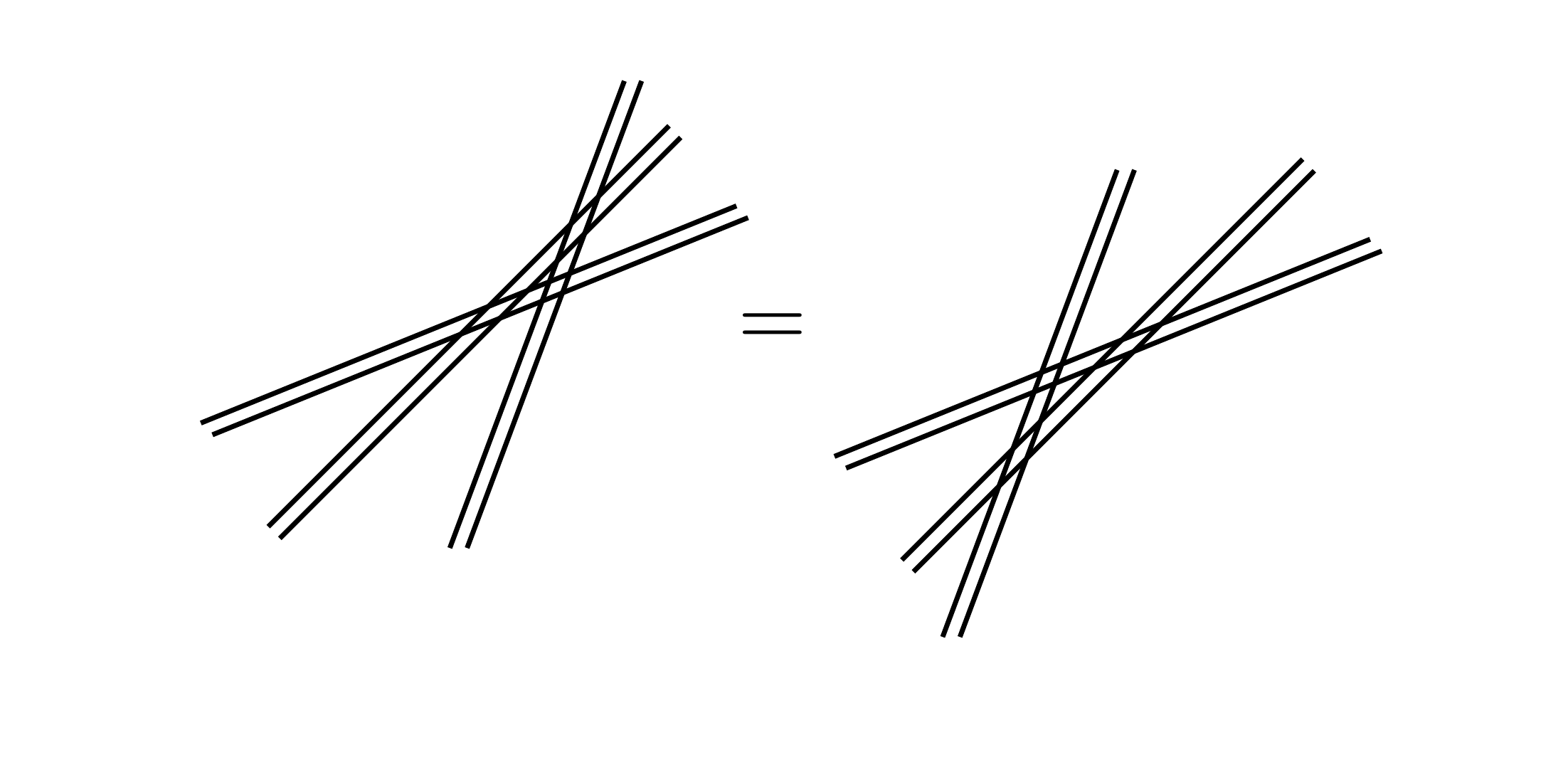}
        \caption{YBE for $\R(u)$ in Eq. (\ref{Rprod})}
		\label{fig:YBE2}
    \end{minipage}
\end{figure}

Another solution $\R(u)$ of the YBE, which maps $({\cal V} \otimes {\cal V}) 
\otimes  ({\cal V} \otimes {\cal V})  \mapsto ({\cal V} 
\otimes {\cal V}) \otimes ({\cal V} \otimes {\cal V})$
is given by the following product of four R-matrices
\be
\R_{12,34}(u) = R_{14}(u)\, R_{13}(u)\, R_{24}(u)\, R_{23}(u)\,,
\label{Rprod}
\ee
which is a $d^{4} \times d^{4}$ matrix. This R-matrix 
can be represented graphically by \emph{two} pairs of lines that cross, as 
shown in Fig. \ref{fig:RR}. The corresponding YBE for $\R$, represented in Fig. 
\ref{fig:YBE2}, follows from the YBE for $R$ shown in Fig. 
\ref{fig:YBE}. A review of models constructed with R-matrices of this 
type can be found in \cite{Zvyagin:2001}.

\begin{figure}[htb]
    \centering
    \begin{minipage}{0.40\textwidth}
        \centering
        \includegraphics[width=0.6\textwidth]{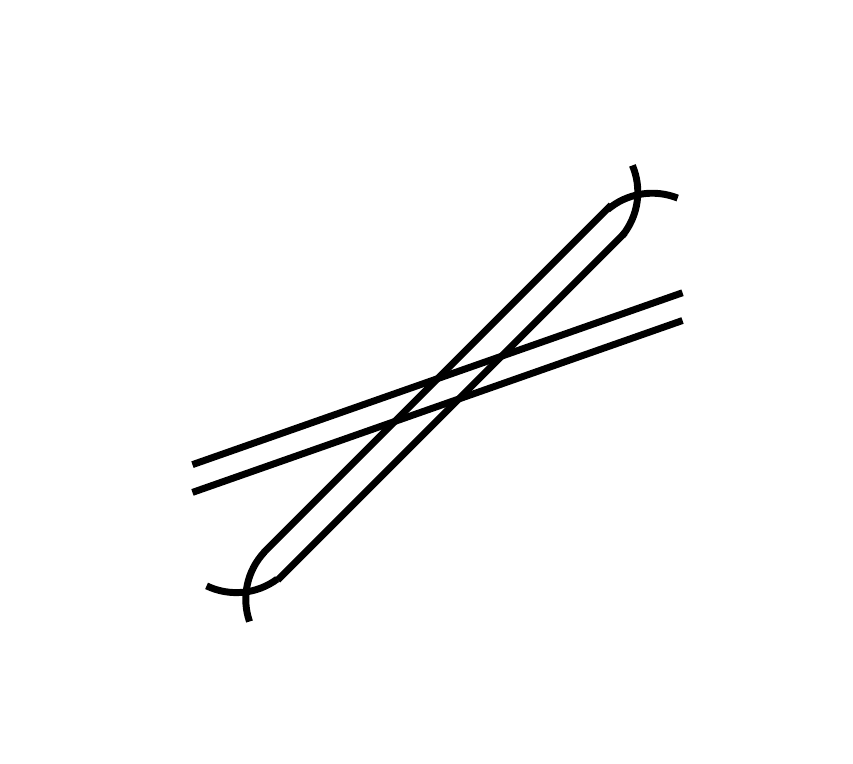} 
        \caption{$\R(u)$ in Eq. (\ref{Rprodgen})}
		\label{fig:RRR}
    \end{minipage}\hfill
    \begin{minipage}{0.40\textwidth}
        \centering
        \includegraphics[width=1.2\textwidth]{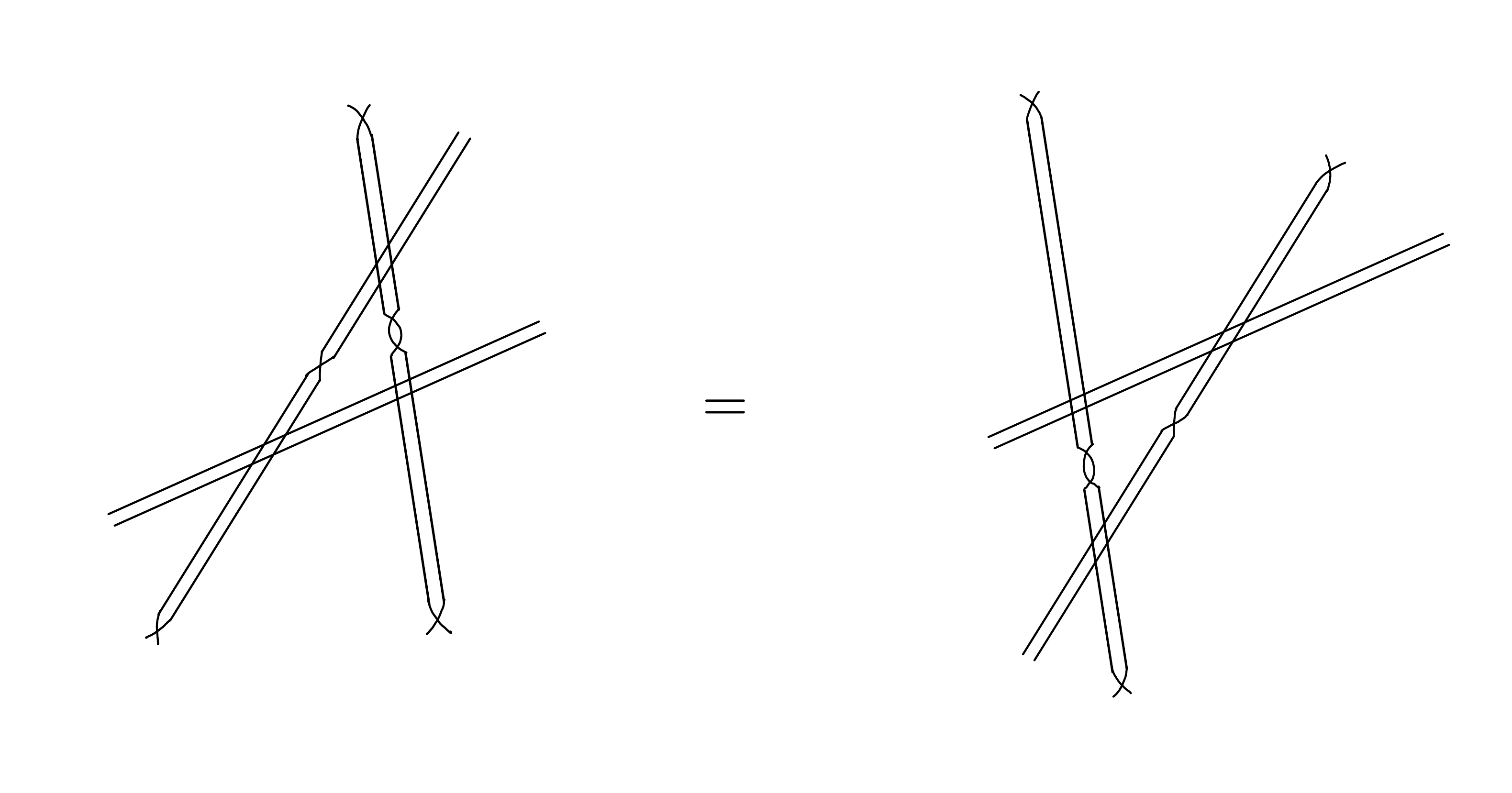}
        \caption{YBE for $\R(u)$ in Eq. (\ref{Rprodgen})}
		\label{fig:YBE3}
    \end{minipage}
\end{figure}

We will need a generalization of the construction (\ref{Rprod}), namely,
\be
\R_{12,34}(u) = R_{43}(-\theta) R_{13}(u)\, R_{14}(u+\theta)\, 
R_{23}(u-\theta)\, R_{24}(u)\, R_{34}(\theta)\,,
\label{Rprodgen}
\ee
where $\theta$ is an arbitrary constant, see Fig. \ref{fig:RRR}.
Indeed, using the regularity 
property (\ref{reg}), the construction (\ref{Rprodgen}) reduces to (\ref{Rprod}) for 
$\theta=0$. The proof that (\ref{Rprodgen}) satisfies the YBE, which 
requires unitarity (\ref{unitarity}) as well as the YBE 
(\ref{YBE}), can also be performed graphically (see Fig. \ref{fig:YBE3}), or by 
a straightforward but long explicit computation.

\subsection{The $D^{(2)}_{2}$ R-matrix}\label{sec:D22}

The $D^{(2)}_{2}$ R-matrix, following a hint
from \cite{Martins1999, Frahm:2012eb}, has recently been shown
\cite{Robertson:2020eri} to be of product form, up to a similarity
transformation.  Indeed, let us write the $D^{(2)}_{2}$ R-matrix 
from \cite{Jimbo:1985ua} as in Appendix A of \cite{Nepomechie:2017hgw}, 
with spectral parameter $u$
and anisotropy parameter $\eta$, and denote it by $\tilde{\R}(u)$.
Then
\be
\tilde{\R}_{12,34}(u) \propto B_{12}\, B_{34}\, \R_{12,34}(u)\, 
B_{12}\, B_{34}\,,
\label{D22R}
\ee
where $\R(u)$ is given by (\ref{Rprodgen}), with $R(u)$ given by the 
$A^{(1)}_{1}$ (XXZ) R-matrix
\be
R(u)=\left(\begin{array}{cccc}
\sinh(-\tfrac{u}{2}+\eta) & 0 & 0 & 0 \\
0 & \sinh(\tfrac{u}{2}) & e^{-\frac{u}{2}} \sinh(\eta) & 0 \\
0 & e^{\frac{u}{2}} \sinh(\eta) & \sinh(\tfrac{u}{2}) & 0 \\
0 & 0 & 0 & \sinh(-\tfrac{u}{2}+\eta) 
\end{array} \right)\,,
\label{Rmat}
\ee
and $\theta = i \pi$. Moreover, the similarity transformation is given 
by
\be
B=\left(\begin{array}{cccc}
1 & 0 & 0 & 0 \\
0 & \frac{\cosh\left(\tfrac{\eta}{2}\right)}{\sqrt{\cosh \eta}} & 
-\frac{\sinh\left(\tfrac{\eta}{2}\right)}{\sqrt{\cosh \eta}}  &  0 \\[0.2cm]
0 & -\frac{\sinh\left(\tfrac{\eta}{2}\right)}{\sqrt{\cosh \eta}} & 
-\frac{\cosh\left(\tfrac{\eta}{2}\right)}{\sqrt{\cosh \eta}} & 0 \\[0.2cm]
0 & 0 & 0 & 1
\end{array} \right),\,\qquad B^{2} = \id\,.
\ee
Following \cite{Saleur:1990vd, Ikhlef:2008zz}, we define the matrix $C$ by
\be
C = \frac{i}{\cosh \eta} {\cal P} R(i \pi)\,, \qquad C^{2} = \id \,.
\label{Cmat}
\ee
Using this notation, the result (\ref{Rprodgen})-(\ref{D22R}) for the $D^{(2)}_{2}$ R-matrix takes the 
final form
\be
\tilde{\R}_{12,34}(u) = B_{12}\, B_{34}\, \R_{12,34}(u)\, B_{12}\, B_{34}\,, 
\label{D22Ra}
\ee
where $\R(u)$ has been redefined (by a simple rescaling) as
\be
\R_{12,34}(u) = 2^{4}\, C_{34}\, R_{14}(u)\, 
R_{13}(u+i \pi)\, R_{24}(u-i\pi)\, R_{23}(u)\, C_{34} \,.
\label{D22Rb}
\ee
Note that we use a tilde to denote similarity-transformed quantities. 
Eqs. (\ref{D22Ra})-(\ref{D22Rb}) are an exact formulation, in our 
notation, of the factorization discovered in \cite{Robertson:2020eri}.
In the isotropic limit $\eta \rightarrow 0$, this result reduces to 
the fact (see e.g. \cite{Gombor:2020kgu}) that the $D_{2}$ (i.e. 
$SO(4)$) R-matrix factorizes into a product of two $A_{1}$ (i.e. 
$SU(2)$) R-matrices, up to a similarity transformation.

For future reference, we note here some useful properties of the 
R-matrix (\ref{Rmat}) in addition to (\ref{YBE})-(\ref{unitarity}): 
quasi-periodicity
\be
R(u+2i\pi) = - R(u)\,,
\label{period}
\ee
PT-symmetry
\be
R_{12}^{t_{1}t_{2}}(u) = R_{21}(u) 
\label{PTsym}
\ee
(where $t_{i}$ denotes transposition in the $i^{th}$ vector space), and 
crossing-unitarity
\be
R_{12}^{t_{1}}(u)\, M_{1}\, R_{12}^{t_{2}}(-u+4\eta)\,  M^{-1}_{1} = 
-\sinh(\tfrac{u}{2})\, \sinh(\tfrac{u}{2}-2\eta)\, \id \,, \qquad 
M=\diag\left(e^{\eta}\,, e^{-\eta} \right) \,.
\label{crossunit}
\ee

\section{The closed $D^{(2)}_{2}$ spin chain}\label{sec:closed}

We begin with the simplest case, namely, the closed periodic
$D^{(2)}_{2}$ spin chain.  In Sec.  \ref{factoridclosed}, 
we use the factorization of the R-matrix
(\ref{D22Ra})-(\ref{D22Rb}) to derive the factorization
identity (\ref{tttrltna})-(\ref{tttrltnb}) that expresses the 
$D^{(2)}_{2}$ transfer matrix as a product of
$A^{(1)}_{1}$ transfer matrices. In Sec.  \ref{ABAclosed}, we use 
this identity to solve the model by means of algebraic Bethe ansatz.

\subsection{Factorization identity}\label{factoridclosed}

The monodromy matrix for a chain of length $N$ is defined by
\be
\tilde{\T}_{0}(u) = \tilde{\R}_{0 N}(u)\, \ldots \tilde{\R}_{0 1}(u) \,,
\label{monodromyTT}
\ee 
where $\tilde{\R}(u)$ is the $D^{(2)}_{2}$ R-matrix.
In order to exploit the factorization (\ref{D22Ra})-(\ref{D22Rb}), it is convenient 
to replace each index $j$ in (\ref{monodromyTT})
(which corresponds to a 4-dimensional vector space)
by a pair of indices $\bar{j}\,, \bar{\bar{j}}$
(each of which corresponds to a 2-dimensional vector space). 
In this way, the monodromy matrix takes the form
\be
\tilde{\T}_{0}(u) = \tilde{\T}_{\bar{0} \bar{\bar{0}}}(u) = 
\tilde{\R}_{\bar{0} 
\bar{\bar{0}}, \bar{N} \bar{\bar{N}}}(u)\, \ldots \tilde{\R}_{\bar{0} 
\bar{\bar{0}}, \bar{1} \bar{\bar{1}}}(u)  \,.
\label{monodromyTTb}
\ee
The relation (\ref{D22Ra}) implies 
\be
\tilde{\T}_{\bar{0} \bar{\bar{0}}}(u) = B_{\bar{0} \bar{\bar{0}}}\, 
\B\, 
\T_{\bar{0} \bar{\bar{0}}}(u)\, 
\B\, 
B_{\bar{0} \bar{\bar{0}}} \,,
\label{TTTrltna}
\ee
where $\T_{\bar{0} \bar{\bar{0}}}(u)$ is defined in terms of $\R$'s as in 
(\ref{monodromyTTb}) except without tildes, and $\B$ is the 
quantum-space operator
\be
\B = B_{\bar{1} \bar{\bar{1}}}\, \ldots 
B_{\bar{N} \bar{\bar{N}}} \,.
\label{Bbb}
\ee
Using (\ref{D22Rb}), we obtain 
\be
\T_{\bar{0} \bar{\bar{0}}}(u) = 2^{4N}\, 
\C\, 
T_{\bar{0}}(u)\, 
T_{\bar{\bar{0}}}(u-i\pi)\, 
\C\,,
\label{TTTrltnb}
\ee
where $T_{\bar{0}}(u)$ is defined by
\be
T_{\bar{0}}(u) = R_{\bar{0} \bar{\bar{N}}}(u)\, R_{\bar{0} 
\bar{N}}(u+i\pi) \ldots R_{\bar{0} \bar{\bar{1}}}(u)\, R_{\bar{0} 
\bar{1}}(u+i\pi) \,,
\label{monodromyT}
\ee 
and $\C$ is the quantum-space operator
\be
\C = C_{\bar{1} \bar{\bar{1}}}\, \ldots C_{\bar{N} \bar{\bar{N}}} \,.
\label{Cbb}
\ee
Note that $T_{\bar{0}}(u)$ is a monodromy matrix on $2N$ sites, with 
$i\pi$ shifts on alternating sites; $T_{\bar{\bar{0}}}(u)$ is given 
by the same expression (\ref{monodromyT}), except with $\bar{0}$ 
replaced by $\bar{\bar{0}}$. Note also the periodicity $T_{\bar{0}}(u+2i\pi) = 
T_{\bar{0}}(u)$ as a consequence of (\ref{period}).

The transfer matrix for the closed periodic spin chain is obtained by 
tracing the monodromy matrix over the auxiliary space
\be
\tilde{\tb}(u) = \tr_{0} \tilde{\T}_{0}(u) 
= \tr_{\bar{0} \bar{\bar{0}}} \tilde{\T}_{\bar{0} \bar{\bar{0}}}(u) \,.
\label{ttdef}
\ee 
Eq. (\ref{TTTrltna}) implies
\be
\tilde{\tb}(u) =   
\B\, 
\tb(u)\,
\B\,, 
\label{tttrltna}
\ee
where $\tb(u)$ is defined in terms of $\T(u)$ as in (\ref{ttdef}) except without tildes.
Using (\ref{TTTrltnb}), we immediately obtain the result
\be
\tb(u) = 2^{4N}\, \C\, 
t(u)\, t(u-i\pi)\, \C \,, 
\label{tttrltnb}
\ee
where $t(u)$ is an $A^{(1)}_{1}$ closed-chain transfer matrix defined by
\be
t(u) = \tr_{\bar{0}} T_{\bar{0}}(u) \,.
\label{transfclosed}
\ee
The result (\ref{tttrltna})-(\ref{tttrltnb}), which we call a 
\emph{factorization identity}, shows that, up to similarity transformations, 
the $D^{(2)}_{2}$ closed-chain transfer matrix is given by a product of 
$A^{(1)}_{1}$ closed-chain transfer matrices with twice as many sites. 

\subsection{Algebraic Bethe ansatz}\label{ABAclosed}

We now proceed to determine the eigenvectors and eigenvalues of the $D^{(2)}_{2}$ 
closed-chain transfer matrix $\tilde{\tb}(u)$ using the factorization 
identity (\ref{tttrltna})-(\ref{tttrltnb}).

To this end, we recall (see e.g. \cite{Faddeev:1996iy})
that the $A^{(1)}_{1}$ transfer matrix can 
be diagonalized by algebraic Bethe ansatz. Indeed, consider the 
general inhomogeneous monodromy matrix with length $L$
\be
T_{0}(u; \{\theta_{l}\}) = R_{0 L}(u-\theta_{L})\, \ldots R_{0 1}(u-\theta_{1}) 
= \left(\begin{array}{cc}
* & {\cal B}(u; \{\theta_{l}\}) \\
* & * 
\end{array} \right)
\,,
\label{monodromyTinhom}
\ee 
where $R(u)$ is given by (\ref{Rmat}), and $\{\theta_{l}\}$ are
arbitrary inhomogeneities. 
(The indices here correspond to 2-dimensional vector spaces, i.e., 
the same as $\bar{j}$ and $\bar{\bar{j}}$ in (\ref{monodromyT}).)
We denote the corresponding closed-chain 
transfer matrix by
\be
t(u; \{\theta_{l}\}) = \tr_{0} T_{0}(u; \{\theta_{l}\}) \,.
\ee
The operator ${\cal B}(u; \{\theta_{l}\})$ in
(\ref{monodromyTinhom}) serves as a creation operator on the reference
state
\be
|0\rangle = {1\choose 0}^{\otimes L} \,.
\label{referencestate}
\ee
The Bethe states defined by
\be
| v_{1} \cdots v_{m} \rangle = \prod_{k=1}^{m} {\cal B}(v_{k}; 
\{\theta_{l}\})\, |0\rangle
\label{Bethestate}
\ee
can be shown to obey the following off-shell equation
\be
t(u; \{\theta_{l}\})\, | v_{1} \cdots v_{m} \rangle = \chi(u; 
\{\theta_{l}\})\, | v_{1} \cdots v_{m} \rangle 
+ \sum_{j=1}^{m} \chi_{j}\, | u, v_{1} \cdots \hat{v}_{j} \cdots 
v_{m} \rangle \,,
\ee
where the variable with a hat is omitted, and
$\chi(u; \{\theta_{l}\})$ is given by
\be
\chi(u; \{\theta_{l}\}) = (-1)^{m}  \left[
\frac{q(u+2\eta)}{q(u)}\prod_{l=1}^{L}\sinh(\eta-\tfrac{1}{2}(u-\theta_{l}))
+\frac{q(u-2\eta)}{q(u)}\prod_{l=1}^{L}\sinh(\tfrac{1}{2}(u-\theta_{l})) 
\right] \,,
\ee
with
\be
q(u) = \prod_{k=1}^{m} \sinh(\tfrac{1}{2}(u-v_{k})) \,.
\ee
Moreover, $\chi_{j}$ is given by
\begin{align}
\chi_{j} &= (-1)^{m+1} \frac{\sinh(\eta)\, e^{\frac{1}{2}(u-v_{j})}}
{\sinh(\tfrac{1}{2}(u-v_{j}))}
\Bigg[ \prod_{l=1}^{L}\sinh(\eta-\tfrac{1}{2}(v_{j}-\theta_{l}))
\prod_{k=1; k \ne j}^{m} 
\frac{\sinh(\tfrac{1}{2}(v_{j}-v_{k})+\eta)}{\sinh(\tfrac{1}{2}(v_{j}-v_{k}))} \non\\
&\qquad - \prod_{l=1}^{L}\sinh(\tfrac{1}{2}(v_{j}-\theta_{l}))
\prod_{k=1; k \ne j}^{m} 
\frac{\sinh(\tfrac{1}{2}(v_{j}-v_{k})-\eta)}{\sinh(\tfrac{1}{2}(v_{j}-v_{k}))} 
\Bigg] \,.
\end{align}

Our original monodromy matrix (\ref{monodromyT}) corresponds to 
setting $L=2N$ in (\ref{monodromyTinhom}), and choosing the 
inhomogeneities as follows
\be
\theta_{l} = \begin{cases}
-i \pi & \text{for $l=$ odd}\\
0 & \text{for $l=$ even}
\end{cases} \,.
\label{inhomogparams}
\ee
It follows that the Bethe states (\ref{Bethestate}) with these 
inhomogeneities are eigenstates of our original transfer matrix (\ref{transfclosed}), with 
corresponding eigenvalues given by
\be
\chi(u) = (-1)^{m} \left(\tfrac{i}{2}\right)^{N}\left[ \sinh^{N}(u-2\eta)\frac{q(u+2\eta)}{q(u)}
+\sinh^{N}(u)\frac{q(u-2\eta)}{q(u)} \right] \,,
\label{chiclosed1}
\ee
provided that $\{v_{k}\}$ satisfy the Bethe equations
\be
\left(\frac{\sinh(v_{j})}{\sinh(v_{j}-2\eta)}\right)^{N} = 
\prod_{k=1; k\ne j}^{m} 
\frac{\sinh(\tfrac{1}{2}(v_{j}-v_{k})+\eta)}{\sinh(\tfrac{1}{2}(v_{j}-v_{k})-\eta)} \,.
\ee
These equations take a symmetric form in terms of $u_{j} \equiv 
v_{j}-\eta$, namely,
\be
\left(\frac{\sinh(u_{j}+\eta)}{\sinh(u_{j}-\eta)}\right)^{N} = 
\prod_{k=1; k\ne j}^{m} 
\frac{\sinh(\tfrac{1}{2}(u_{j}-u_{k})+\eta)}{\sinh(\tfrac{1}{2}(u_{j}-u_{k})-\eta)} \,.
\label{BEclosed}
\ee
Setting
\be
Q(u) = \prod_{k=1}^{m} \sinh(\tfrac{1}{2}(u-u_{k})) = q(u+\eta) \,,
\ee
the expression for the eigenvalues (\ref{chiclosed1}) of the 
$A^{(1)}_{1}$ closed-chain transfer matrix $t(u)$ (\ref{transfclosed}) take the final form
\be
\chi(u) = (-1)^{m} \left(\tfrac{i}{2}\right)^{N} \left[ \sinh^{N}(u-2\eta)\frac{Q(u+\eta)}{Q(u-\eta)}
+\sinh^{N}(u)\frac{Q(u-3\eta)}{Q(u-\eta)} \right] \,.
\label{chiclosed2}
\ee

Coming back to the $D^{(2)}_{2}$ closed-chain transfer matrix 
$\tilde{\tb}(u)$ (\ref{ttdef}), we conclude from the 
factorization identity (\ref{tttrltna})-(\ref{tttrltnb}) 
that its Bethe states are given by 
\be
\B\, \C\, | v_{1} 
\cdots v_{m} \rangle \,,
\ee
where the vectors $| v_{1} \cdots v_{m} \rangle$ are given by (\ref{Bethestate}),
and $\B$ and $\C$ are given respectively by (\ref{Bbb}) and 
(\ref{Cbb}),
see \cite{Martins1999} for an alternative approach. Moreover, 
the corresponding eigenvalues $\Lambda(u)$ are given by
\be
\Lambda(u) = 2^{4N}\, \chi(u)\, \chi(u-i\pi) \,,
\ee
where $\chi(u)$ is given by (\ref{chiclosed2}), and the associated Bethe 
equations are given by (\ref{BEclosed}). The latter results agree with 
expressions obtained by Reshetikhin using analytical Bethe ansatz 
\cite{Reshetikhin:1987}.

\subsection{$Z_{2}$ symmetry}\label{sec:Z2closed}

The transfer matrix $t(u)$ (\ref{transfclosed}) has the property
\be
\C\, t(u)\, 
\C
= (-1)^N \, t(u+ i\pi) \,,
\label{Ctclosed}
\ee
where $\C$ (\ref{Cbb}) is defined in terms of $C$ (\ref{Cmat}).
The proof is short: the fact that the R-matrix satisfies the identity
\be
C_{23}\, R_{13}(u)\, R_{12}(u+i\pi)\, C_{23} =  R_{13}(u+i\pi)\, 
R_{12}(u) 
\ee
implies that the monodromy matrix (\ref{monodromyT}) satisfies the
corresponding identity
\be
\C\, T_{\bar{0}}(u)\, 
\C
= (-1)^N \, T_{\bar{0}}(u+ i\pi) \,. 
\label{Tdiscrete}
\ee
By tracing over the auxiliary space $\bar{0}$, we obtain (\ref{Ctclosed}).

The property (\ref{Ctclosed}) implies that the $D^{(2)}_{2}$ transfer matrix 
$\tb(u)$ (\ref{tttrltnb}) can also be written in the form
\be
\tb(u) = 2^{4N}\, 
t(u+i\pi)\, t(u) \,,
\ee
and therefore it has the $Z_{2}$ symmetry
\be
\C\, \tb(u)\,
\C
= \tb(u) \,.
\label{Csymmetryclosed}
\ee
The $Z_{2}$ symmetry of the staggered six-vertex model
was noted already in \cite{Ikhlef:2008zz}.

\subsection{Degeneracies}\label{sec:degclosed}

For real values of $\eta$, each of the eigenvalues of $t(u)$
(\ref{transfclosed}) is either a singlet or a doublet (2-fold
degenerate).  However, as the result of the $Z_{2}$ symmetry, some of
the degeneracies of $\tb(u)$ (\ref{tttrltnb}) become \emph{doubled}, 
leading to doublets or quartets.

The key point is that the $Z_{2}$ symmetry shifts the
argument of the ${\cal B}$-operator by $i\pi$
\be
\C\, 
{\cal B}(u)\,
\C
= (-1)^N \, {\cal B}(u+ i\pi) \,,
\label{Bshiftclosed}
\ee 
as follows from (\ref{monodromyTinhom}) and
(\ref{Tdiscrete}). The Bethe states (\ref{Bethestate}) therefore transform as follows
\be
\C\,
| v_{1} \cdots v_{m} \rangle =  (-1)^{N\, m}\, | v_{1}+ i\pi \cdots v_{m}+ i\pi 
\rangle \,,
\ee
since the reference state remains invariant $\C\, |0\rangle = 
|0\rangle$. In other words, under the $Z_{2}$ symmetry, each of the 
Bethe roots $v_{k}$ (or, equivalently, $u_{k}$) is shifted by 
$i\pi$.  If $Q(u+ i\pi)\ne \pm Q(u)$, then the Bethe states 
corresponding to $Q(u)$ and $Q(u+ i\pi)$ are mapped into each 
other by the $Z_{2}$ symmetry $\C$. (The argument is the same as for 
the open chain, which is presented in Sec. \ref{sec:degopen}.) 
It follows from (\ref{Csymmetryclosed}) 
that the two Bethe states have the same eigenvalue of 
$\tb(u)$, which means that they are degenerate.

Our goal in the remainder of this paper is to obtain factorization 
identities analogous to (\ref{tttrltna})-(\ref{tttrltnb}) for $D^{(2)}_{2}$ open-chain transfer 
matrices, and use these relations to solve the models.

\section{The open $D^{(2)}_{2}$ spin chain}\label{sec:open}

We turn now to the open $D^{(2)}_{2}$ spin chain.  We will consider
two different sets of integrable boundary conditions, corresponding to
the two possible values (namely, 0 and 1) of a certain parameter $\varepsilon$.  
As before, our strategy will be to use factorization identities to solve 
the models. After introducing the transfer matrix in Sec. 
\ref{sec:transfer}, we consider the case $\varepsilon=1$ in 
Sec. \ref{sec:eps1}, followed by case $\varepsilon=0$ in Sec. 
\ref{sec:eps0}.

\subsection{Transfer matrix}\label{sec:transfer}

In order to construct an integrable open-chain transfer matrix
\cite{Sklyanin:1988yz}, we need not only an R-matrix, but also a
K-matrix, i.e., a solution of the corresponding boundary Yang-Baxter
equation \cite{Sklyanin:1988yz, Cherednik:1985vs, Ghoshal:1993tm}. For $D^{(2)}_{n+1}$, such K-matrices have been found in 
\cite{Martins:2000xie, Nepomechie:2018wzp}. The K-matrices in 
\cite{Nepomechie:2018wzp} depend on two discrete parameters:  $p$ (which can take $n+1$ different 
values, namely, $p=0, 1, \ldots, n$) and $\varepsilon$ (which can take 
two different values, namely, $\varepsilon = 0, 1$). We consider here $n=1$ 
(corresponding to $D^{(2)}_{2}$); and, for concreteness, we set 
$p=0$. (The case $p=1$ is simply related to the case $p=0$ by a $p \leftrightarrow 
n-p$ duality symmetry \cite{Nepomechie:2018wzp, Nepomechie:2019tbr}.) The right K-matrix, which we denote 
here by $\tilde{\K}^{R}(u)$, is then given by
\be 
\tilde{\K}^{R}(u) =  
\left( \begin{array}{cccc}
g(u) & 0 & 0 & 0 \\
0 & k_{1}(u) & k_{2}(u) & 0 \\
0 & k_{2}(u) & k_{1}(u) & 0 \\
0 & 0 & 0 & g(u)
\end{array} \right) \,,
\label{KR}
\ee
where
\begin{align}
g(u) &= \frac{\cosh(u-\eta + 
\frac{i\pi}{2}\varepsilon)}{\cosh(u+\eta - \frac{i\pi}{2}\varepsilon)} \,, \non \\
k_{1}(u)  &= \frac{\cosh(u) \cosh(\eta + \frac{i\pi}{2}\varepsilon)}
{\cosh(u+\eta+\frac{i\pi}{2}\varepsilon)} 
\,, \non \\
k_{2}(u)  &= -\frac{\sinh(u) \sinh(\eta + \frac{i\pi}{2}\varepsilon)}
{\cosh(u+\eta + \frac{i\pi}{2}\varepsilon)} \,,
\label{functions}
\end{align}
with $\varepsilon = 0, 1$. For the left K-matrix, we take \cite{Nepomechie:2018wzp}
\be
\tilde{\K}^{L}(u) = \tilde{\K}^{R}(-u+2\eta)\, \M \,, \qquad \M = M \otimes M \,,
\label{KL}
\ee 
where $M$ is defined in (\ref{crossunit}), 
so that the transfer matrix has quantum-group symmetry, see Sec. 
\ref{sec:sym}.

The $D^{(2)}_{2}$ open-chain transfer matrix for a chain with $N$ sites is given by \cite{Sklyanin:1988yz}
\be
\tilde{\tb}(u) = \tr_{0} \Big\{  \tilde{\K}^{L}_{0}(u)\, \tilde{\T}_{0}(u)\,  
\tilde{\K}^{R}_{0}(u)\, 
\widehat{\tilde{\T}}_{0}(u) \Big\} \,, 
\label{ttopen}
\ee
where $\tilde{\T}_{0}(u)$ is given by (\ref{monodromyTT}) and 
(\ref{monodromyTTb}). Similarly, $\widehat{\tilde{\T}}_{0}(u)$ is given by
\be
\widehat{\tilde{\T}}_{0}(u) = \tilde{\R}_{1 0}(u)\, \ldots 
\tilde{\R}_{N 0}(u) \,,
\label{hatmonodromyTT}
\ee
or equivalently
\be
\widehat{\tilde{\T}}_{\bar{0} \bar{\bar{0}}}(u) 
= \tilde{\R}_{\bar{1} \bar{\bar{1}}, \bar{0} \bar{\bar{0}}}(u) \ldots
\tilde{\R}_{\bar{N} \bar{\bar{N}}, \bar{0} \bar{\bar{0}}}(u) \,,
\label{hatmonodromyTTb}
\ee
where we have replaced (as we did for $\tilde{\T}_{0}(u)$ in  Sec. \ref{factoridclosed}) each index 
$j$ in (\ref{hatmonodromyTT}) by a pair of indices $\bar{j}\,, \bar{\bar{j}}$.  
Eq. (\ref{D22Ra}) then implies
\be
\widehat{\tilde{\T}}_{\bar{0} \bar{\bar{0}}}(u) = B_{\bar{0} \bar{\bar{0}}}\, 
\B\, 
\widehat{\T}_{\bar{0} \bar{\bar{0}}}(u)\, 
\B\, 
B_{\bar{0} \bar{\bar{0}}} \,,
\label{hatTTTrltna}
\ee
where $\widehat{\T}_{\bar{0} \bar{\bar{0}}}(u)$ is defined in terms of $\R$'s as in 
(\ref{hatmonodromyTTb}) except without tildes. Using (\ref{D22Rb}), we 
obtain
\be
\widehat{\T}_{\bar{0} \bar{\bar{0}}}(u) = 2^{4N}\,   
C_{\bar{0} \bar{\bar{0}}}\, 
\widehat{T}_{\bar{\bar{0}}}(u+i\pi)\, 
\widehat{T}_{\bar{0}}(u)\, 
C_{\bar{0} \bar{\bar{0}}}\,, 
\label{hatTTTrltnb}
\ee
where $\widehat{T}_{\bar{0}}(u)$ is defined by
\be
\widehat{T}_{\bar{0}}(u) = R_{\bar{1} \bar{0} }(u-i\pi)\, 
R_{\bar{\bar{1}} \bar{0}}(u)\, \ldots
R_{\bar{N} \bar{0}}(u-i\pi)\,
R_{\bar{\bar{N}} \bar{0}}(u) \,,
\label{hatmonodromyT}
\ee 
and $\widehat{T}_{\bar{\bar{0}}}(u)$ is given 
by the same expression (\ref{hatmonodromyT}), except with $\bar{0}$ 
replaced by $\bar{\bar{0}}$.

\subsection{The case $\varepsilon=1$}\label{sec:eps1} 

For the case $\varepsilon=1$, the transfer matrix $\tilde{\tb}(u)$ (\ref{ttopen}) 
satisfies 
\be
\tilde{\tb}(u) = \B\, 
\tb(u)\,
\B \,,
\label{tttrltneps1a}
\ee
where $\tb(u)$ satisfies the remarkable factorization identity
\be
\tb(u) = \phi(u)\, t(u+i\pi)\, t(u) \,, 
\qquad \phi(u) = 2^{8N}\frac{\sinh u\, 
\sinh(u-2\eta)}{\sinh(u+\eta)\sinh(u-3\eta)} \,,
\label{tttrltneps1b}
\ee
where $t(u)$ is an $A^{(1)}_{1}$ open-chain transfer matrix defined by
\be
t(u) = \tr_{\bar{0}} \Big\{ 
M_{\bar{0}}\, 
T_{\bar{0}}(u)\, 
\widehat{T}_{\bar{0}}(u+ i\pi)\Big\}\,,
\label{teps1}
\ee
and $T_{\bar{0}}(u)$ and $\widehat{T}_{\bar{0}}(u)$
are defined in (\ref{monodromyT}) and (\ref{hatmonodromyT}), respectively.
The proof of this factorization identity is presented in Appendix 
\ref{sec:factopen}. 
Note the periodicity $t(u + 2i\pi) = t(u)$ as a consequence of 
(\ref{period}).

Notice the shift by $i\pi$ in the argument of $\widehat{T}$ (compared
with $T$) in the transfer matrix (\ref{teps1}).  While this shift may
appear innocuous, its effects are profound.  To our knowledge,
open-chain transfer matrices with such shifts have not been
considered before; a priori, it is not even clear whether such transfer
matrices commute for different values of the spectral parameter.  

We will interpret such a shift as the rapidity of the boundary; or 
equivalently, as a boundary inhomogeneity.
We will then proceed to diagonalize the
transfer matrix.

\subsubsection{Transfer matrix with a moving 
boundary}\label{sec:moving}

As in the closed-chain case (see (\ref{monodromyTinhom})), it is convenient to consider 
a slightly more general problem, namely, a chain of length $L$ with
arbitrary inhomogeneities at each site. The monodromy matrices are therefore given by
\begin{align}
T_{0}(u; \{\theta_{l}\}) &= R_{0 L}(u-\theta_{L})\, \ldots R_{0 
1}(u-\theta_{1}) \,, \non\\
\widehat{T}_{0}(u; \{\theta_{l}\}) &= R_{1 0}(u+\theta_{1})\, \ldots 
R_{L 0}(u+\theta_{L}) \,,
\label{monodromyTinhom2}
\end{align}
where $R(u)$ is given by (\ref{Rmat}), and $\{\theta_{l}\}$ are 
arbitrary inhomogeneities,
cf. (\ref{monodromyT}) and (\ref{hatmonodromyT}). These monodromy 
matrices satisfy the familiar fundamental relations
\begin{align}
R_{0 0'}(u-v)\, T_{0}(u; \{\theta_{l}\})\, 
T_{0'}(v; \{\theta_{l}\}) 
&=
T_{0'}(v; \{\theta_{l}\})\,
T_{0}(u; \{\theta_{l}\})\, 
R_{0 0'}(u-v) \,, \non\\
R_{0 0'}(u-v)\, \widehat{T}_{0'}(u; \{\theta_{l}\})\, 
\widehat{T}_{0}(v; \{\theta_{l}\}) 
&=
\widehat{T}_{0}(v; \{\theta_{l}\})\,
\widehat{T}_{0'}(u; \{\theta_{l}\})\, 
R_{0 0'}(u-v) \,, \non\\
T_{0}(u; \{\theta_{l}\})\,  
R_{0 0'}(u+v)\,
\widehat{T}_{0'}(v; \{\theta_{l}\})\, 
&=
\widehat{T}_{0'}(v; \{\theta_{l}\})\, 
R_{0 0'}(u+v)\,
T_{0}(u; \{\theta_{l}\})\,.
\label{fundamrltns}
\end{align}
Moreover, we 
consider the transfer matrix
\be
t(u; \{\theta_{l}\}) = \tr_{0} \Big\{  M_{0}\, {\cal U}_{0}(u; 
\{\theta_{l}\}) \Big\} \,, 
\qquad {\cal U}_{0}(u; \{\theta_{l}\}) = T_{0}(u; \{\theta_{l}\})\, 
\widehat{T}_{0}(u+u_{0}; \{\theta_{l}\}) \,,
\label{topen}
\ee
where the shift $u_{0}$ in the argument of $\widehat{T}$ is arbitrary. The transfer matrix for our problem 
(\ref{teps1}) is clearly a special case of 
(\ref{topen}).\footnote{Although not necessary here, we note that
it is possible to further generalize the transfer matrix (\ref{topen}) 
by introducing general K-matrices, namely
\be
t(u; \{\theta_{l}\}) = \tr_{0} \Big\{  K^{L}_{0}(u)\, {\cal U}_{0}(u; 
\{\theta_{l}\}) \Big\} \,, 
\qquad {\cal U}_{0}(u; \{\theta_{l}\}) = T_{0}(u; \{\theta_{l}\})\, K^{R}_{0}(u)\, 
\widehat{T}_{0}(u+u_{0}; \{\theta_{l}\}) \,, \non
\ee
where $K^{R}(u)$ satisfies the BYBE (\ref{BYBE}), i.e.
\be
R_{12}(u-v)\, K^{R}_{1}(u)\, R_{21}(u+v+u_{0})\, 
K^{R}_{2}(v) = 
K^{R}_{2}(v)\, R_{12}(u+v+u_{0})\, K^{R}_{1}(u)\, R_{21}(u-v) \,. \non
\ee
This equation has the solution 
$K^{R}(u)=\id$ if $\left[\check{R}(u) \,, \check{R}(v) \right] = 0$. 
Moreover, in order to ensure the commutativity (\ref{commutativity}),
$K^{L}(u)$ satisfies
\begin{align}
	& R_{12}(v-u)\, K^{L\, t_{1}}_{1}(u)\, M^{-1}_{1}\, 
	R_{12}^{t_{12}}(-u-v-u_{0}+4\eta)\, M_{1}\, 
	K^{L\, t_{2}}_{2}(v) \non \\
	&\quad = K^{L\, t_{2}}_{2}(v)\, M_{1}\, 
	R_{12}(-u-v-u_{0}+4\eta)\, M^{-1}_{1}\, 
	K^{L\, t_{1}}_{1}(u)\, R_{12}^{t_{12}}(v-u) \,. \non 
\end{align}
This equation has the solution $K^{L}(u) = K^{R}(-u -u_{0} + 2\eta)\, 
M$ if $K^{R}(u)$ satisfies (\ref{BYBE}).

We also note that (\ref{BYBE}) can be mapped to the usual BYBE by 
performing the shifts $u \mapsto u - u_{0}/2$ and $v \mapsto v - 
u_{0}/2$. Hence, the above $K^{R}(u)$ can be constructed from a 
solution of the usual BYBE by shifting the rapidity by $u_{0}/2$.
\label{foot}}

It is straightforward to show using (\ref{fundamrltns})
and $\left[\check{R}(u) \,, \check{R}(v) \right] = 0$ (where 
$\check{R}(u) \equiv {\cal P} R(u)$), 
that the double-row monodromy matrix
${\cal U}(u; \{\theta_{l}\})$ (\ref{topen}) obeys the following boundary Yang-Baxter 
equation (BYBE) 
\begin{align}
& R_{12}(u-v)\, {\cal U}_{1}(u; \{\theta_{l}\})\, R_{21}(u+v+u_{0})\, 
{\cal U}_{2}(v; \{\theta_{l}\}) \non\\
&\qquad = 
{\cal U}_{2}(v; \{\theta_{l}\})\, R_{12}(u+v+u_{0})\, {\cal U}_{1}(u; 
\{\theta_{l}\})\, R_{21}(u-v) \,.
\label{BYBE}
\end{align}
Note the shift by $u_{0}$ in the R-matrix whose argument has the sum
of rapidities.  It implies that if a ``particle'' approaches the
boundary with rapidity $u$, then after reflection the particle has
rapidity $-u-u_{0}$.  We can attribute this shift
to a moving boundary, with rapidity $u_{0}$. Equivalently, this shift can be regarded as 
a boundary inhomogeneity, as opposed to the bulk inhomogeneities $\{ 
\theta_{l} \}$.

Despite the presence of a shift in the BYBE, the transfer matrix 
nevertheless has the crucial commutativity property
\be
\left[t(u; \{\theta_{l}\}) \,, t(v; \{\theta_{l}\}) \right] = 0 \,.
\label{commutativity}
\ee
Indeed, the commutativity proof in \cite{Sklyanin:1988yz} can be readily 
generalized to accommodate this shift, for arbitrary values of $u_{0}$.

\subsubsection{Algebraic Bethe ansatz}\label{sec:ABAeps1} 

We now proceed to diagonalize the transfer matrix (\ref{topen}) by
algebraic Bethe ansatz. Following \cite{Sklyanin:1988yz}, we set
\be
{\cal U}_{0}(u; \{\theta_{l}\}) = \left(\begin{array}{cc}
* & {\cal B}(u; \{\theta_{l}\}) \\
* & * 
\end{array} \right) \,,
\label{Boperator}
\ee
and act with ${\cal B}(u; \{\theta_{l}\})$ on the reference state 
(\ref{referencestate}) to create the Bethe states
\be
| v_{1} \cdots v_{m} \rangle = \prod_{k=1}^{m} {\cal B}(v_{k}; 
\{\theta_{l}\})\, |0\rangle \,,
\label{Bethestateopen}
\ee
which obey the following off-shell equation
\be
t(u; \{\theta_{l}\})\, | v_{1} \cdots v_{m} \rangle = \chi(u; 
\{\theta_{l}\})\, | v_{1} \cdots v_{m} \rangle 
+ \sum_{j=1}^{m} \chi_{j}\, | u, v_{1} \cdots \hat{v}_{j} \cdots 
v_{m} \rangle \,.
\label{offshellopen}
\ee
Here, $\chi(u; \{\theta_{l}\})$ is given by
\begin{align}
\chi(u; \{\theta_{l}\}) &=  
\frac{\sinh(u+\tfrac{u_{0}}{2}-2\eta)}{\sinh(u+\tfrac{u_{0}}{2}-\eta)}
\frac{q(u+2\eta)}{q(u)}\prod_{l=1}^{L}\sinh(\tfrac{1}{2}(u-\theta_{l})-\eta)\, 
\sinh(\tfrac{1}{2}(u+u_{0}+\theta_{l})-\eta)\non\\
&\qquad
+\frac{\sinh(u+\tfrac{u_{0}}{2})}{\sinh(u+\tfrac{u_{0}}{2}-\eta)}
\frac{q(u-2\eta)}{q(u)}\prod_{l=1}^{L}\sinh(\tfrac{1}{2}(u-\theta_{l}))\, \sinh(\tfrac{1}{2}(u+u_{0}+\theta_{l})) 
 \,,
\end{align}
with
\be
q(u) = \prod_{k=1}^{m} \sinh(\tfrac{1}{2}(u-v_{k}))\, 
\sinh(\tfrac{1}{2}(u+u_{0}+v_{k})-\eta) \,,
\label{qfunc}
\ee
and $\chi_{j}$ is given by
\begin{align}
\chi_{j} &= - \frac{\sinh(\eta)\, \sinh(u+\tfrac{u_{0}}{2}-2\eta)}
{\sinh(\tfrac{1}{2}(u-v_{j}))\, \sinh(\tfrac{1}{2}(u+u_{0}+v_{j})-\eta)}
\frac{\sinh(v_{j}+\tfrac{u_{0}}{2})}{\sinh(v_{j}+\tfrac{u_{0}}{2}-\eta)} \non\\
&\quad \times
\Bigg[ \prod_{l=1}^{L}\sinh(\tfrac{1}{2}(v_{j}-\theta_{l})-\eta)\, 
\sinh(\tfrac{1}{2}(v_{j}+ u_{0}+\theta_{l})-\eta) \non\\
&\qquad\times
\prod_{k=1; k \ne j}^{m} 
\frac{\sinh(\tfrac{1}{2}(v_{j}-v_{k})+\eta)\, 
\sinh(\tfrac{1}{2}(v_{j}+v_{k}+u_{0}))}
{\sinh(\tfrac{1}{2}(v_{j}-v_{k}))\, 
\sinh(\tfrac{1}{2}(v_{j}+v_{k}+u_{0})-\eta)} \non\\
&\quad 
- \prod_{l=1}^{L}\sinh(\tfrac{1}{2}(v_{j}-\theta_{l}))\, 
\sinh(\tfrac{1}{2}(v_{j}+ u_{0}+\theta_{l})) \non\\
&\qquad\times \prod_{k=1; k \ne j}^{m} 
\frac{\sinh(\tfrac{1}{2}(v_{j}-v_{k})-\eta)\, 
\sinh(\tfrac{1}{2}(v_{j}+v_{k}+u_{0})-2\eta)}
{\sinh(\tfrac{1}{2}(v_{j}-v_{k}))\, \sinh(\tfrac{1}{2}(v_{j}+v_{k}+u_{0})-\eta)} 
\Bigg] \,.
\label{chijopen}
\end{align}
Note that a nonzero value of $u_{0}$ indeed profoundly affects the solution.

Our original monodromy matrices (\ref{monodromyT}) and (\ref{hatmonodromyT}) correspond to 
setting $L=2N$ in (\ref{monodromyTinhom2}), and choosing the 
inhomogeneities $\{\theta_{l}\}$ as in (\ref{inhomogparams}). Moreover, our original transfer matrix (\ref{teps1})
corresponds to setting the shift $u_{0}= i \pi$ in (\ref{topen}).
It follows that the 
Bethe states (\ref{Bethestateopen}) with these 
parameter values are eigenstates of our original transfer matrix (\ref{teps1}), with 
corresponding eigenvalues given by
\be
\chi(u) =  
\left(-\tfrac{1}{4}\right)^{N} \left[
\frac{\cosh(u-2\eta)}{\cosh(u-\eta)}\frac{q(u+2\eta)}{q(u)}\sinh^{2N}(u-2\eta)
+\frac{\cosh(u)}{\cosh(u-\eta)}\frac{q(u-2\eta)}{q(u)}\sinh^{2N}(u)\right]
\label{chiopeneps1a}
\ee
with
\be
q(u) = \prod_{k=1}^{m} \sinh(\tfrac{1}{2}(u-v_{k}))\, 
\cosh(\tfrac{1}{2}(u+v_{k})-\eta) \,,
\ee
provided that $\{v_{k}\}$ satisfy the Bethe equations
\be
\left(\frac{\sinh(v_{j})}{\sinh(v_{j}-2\eta)}\right)^{2N} = 
\prod_{k=1; k\ne j}^{m} 
\frac{\sinh(\tfrac{1}{2}(v_{j}-v_{k})+\eta)\, 
\cosh(\tfrac{1}{2}(v_{j}+v_{k}))}
{\sinh(\tfrac{1}{2}(v_{j}-v_{k})-\eta)\, 
\cosh(\tfrac{1}{2}(v_{j}+v_{k})-2\eta)} \,.
\ee
These equations take a symmetric form in terms of $u_{j} \equiv 
v_{j}-\eta$, namely,
\be
\left(\frac{\sinh(u_{j}+\eta)}{\sinh(u_{j}-\eta)}\right)^{2N} = 
\prod_{k=1; k\ne j}^{m} 
\frac{\sinh(\tfrac{1}{2}(u_{j}-u_{k})+\eta)\, 
\cosh(\tfrac{1}{2}(u_{j}+u_{k})+\eta)\,}
{\sinh(\tfrac{1}{2}(u_{j}-u_{k})-\eta)\, 
\cosh(\tfrac{1}{2}(u_{j}+u_{k})-\eta)} \,.
\label{BEopeneps1}
\ee
Setting
\be
Q(u) = \prod_{k=1}^{m} \sinh(\tfrac{1}{2}(u-u_{k}))\, 
\cosh(\tfrac{1}{2}(u+u_{k})) = q(u+\eta) \,,
\label{Qeps1}
\ee
the expression for the eigenvalues (\ref{chiopeneps1a}) of the 
$A^{(1)}_{1}$ open-chain transfer matrix $t(u)$ (\ref{teps1}) take the final form
\be
\chi(u) =  
\left(-\tfrac{1}{4}\right)^{N} \left[
\frac{\cosh(u-2\eta)}{\cosh(u-\eta)}\frac{Q(u+\eta)}{Q(u-\eta)}\sinh^{2N}(u-2\eta)
+\frac{\cosh(u)}{\cosh(u-\eta)}\frac{Q(u-3\eta)}{Q(u-\eta)}\sinh^{2N}(u)\right] \,.
\label{chiopeneps1b}
\ee
Returning to the $D^{(2)}_{2}$ open-chain transfer matrix 
$\tilde{\tb}(u)$ (\ref{ttopen}) with $\varepsilon=1$, we conclude from the 
factorization identity (\ref{tttrltneps1a})-(\ref{tttrltneps1b}) that its Bethe states are given by 
\be
\B\,  | v_{1} \cdots v_{m} \rangle \,,
\ee
where the vectors $| v_{1} \cdots v_{m} \rangle$ are given by (\ref{Bethestateopen}),
and $\B$ is given by (\ref{Bbb}),
which is a new result. Moreover, 
the corresponding eigenvalues $\Lambda(u)$ are given by
\be
\Lambda(u) = \phi(u)\, \chi(u)\, \chi(u+i\pi) \,,
\ee
where $\chi(u)$ is given by (\ref{chiopeneps1b}), and the associated Bethe 
equations are given by (\ref{BEopeneps1}). The latter results agree with 
the recent proposal in \cite{Robertson:2020imc}, which improved on an earlier proposal \cite{Nepomechie:2019tbr}.

\subsubsection{Symmetries}\label{sec:sym}

We briefly discuss here the quantum group (QG) and 
$Z_{2}$ symmetries of the transfer matrix, which we will then use to understand 
the degeneracies of the spectrum.

\noindent
{\bf Quantum group symmetry}

The $D^{(2)}_{2}$ open-chain transfer matrix $\tilde{\tb}(u)$ 
(\ref{ttopen}) has the QG symmetry $U_{q}(B_{1})$  
\cite{Nepomechie:2017hgw,Nepomechie:2018wzp}
\be
\left[ \Delta_{N}(\tilde{\Hb}) \,, \tilde{\tb}(u) \right] = 0 \,, \qquad
\left[ \Delta_{N}(\tilde{\E}^{\pm}) \,, \tilde{\tb}(u) \right] = 0 \,, 
\ee
where the generators at one site are given by
\be
\tilde{\Hb} = \diag\left(1, 0, 0, -1 \right)\,, \qquad
\tilde{\E}^{+} = \frac{1}{\sqrt{2}}
\left( 
\begin{array}{cccc}
0 & 1 & 1 & 0 \\
0 & 0 & 0 & -1 \\
0 & 0 & 0 & -1 \\
0 & 0 & 0 & 0
\end{array}\right)\,, \qquad \tilde{\E}^{-} =\tilde{\E}^{+\ t} \,,
\ee
and the two-site coproducts are given by
\begin{align}
\Delta(\tilde{\Hb}) &= \tilde{\Hb} \otimes \id + \id\otimes 
\tilde{\Hb}  \,, \non \\
\Delta(\tilde{\E}^{\pm}) &= \tilde{\E}^{\pm} \otimes e^{\eta\, \tilde{\Hb}} +
e^{-\eta\, \tilde{\Hb}}\otimes \tilde{\E}^{\pm} \,.
\end{align}
Higher coproducts follow, as usual, from coassociativity 
$\left( \Delta \otimes \id \right)\Delta = \left( \id \otimes \Delta \right)\Delta$.
These generators satisfy 
\be
\left[\Delta(\tilde{\Hb}) \,, \Delta(\tilde{\E}^{\pm}) \right] = \pm 
\Delta(\tilde{\E}^{\pm}) \,, \qquad
\left[\Delta(\tilde{\E}^{+}) \,, \Delta(\tilde{\E}^{-}) \right] = 
\frac{\sinh(2\eta \Delta(\tilde{\Hb}))}{\sinh(2\eta)} \,.
\ee
Performing the (inverse) similarity transformation, we obtain
\begin{align}
\Hb &= B\, \tilde{\Hb}\, B = \diag\left(1, 0, 0, -1 \right) \,, \non\\
\E^{+} &= B\, \tilde{\E}^{+} \, B = \frac{1}{\sqrt{2\cosh \eta}}
\left( 
\begin{array}{cccc}
0 & e^{-\frac{\eta}{2}} & -e^{\frac{\eta}{2}} & 0 \\
0 & 0 & 0 & -e^{-\frac{\eta}{2}} \\
0 & 0 & 0 & e^{\frac{\eta}{2}} \\
0 & 0 & 0 & 0
\end{array}\right)\,, \qquad 
\E^{-} = B\, \tilde{\E}^{-} \, B = \E^{+\ t} \,,
\label{QGgens}
\end{align}
and
\begin{align}
\Delta(\Hb) &= \Hb \otimes \id + \id\otimes 
\Hb  \,, \non \\
\Delta(\E^{\pm}) &= \E^{\pm} \otimes e^{\eta\, \Hb} +
e^{-\eta\, \Hb}\otimes \E^{\pm} \,,
\label{coproductswithouttilde}
\end{align}
with
\be
\left[\Delta(\Hb) \,, \Delta(\E^{\pm}) \right] = \pm 
\Delta(\E^{\pm}) \,, \qquad
\left[\Delta(\E^{+}) \,, \Delta(\E^{-}) \right] = 
\frac{\sinh(2\eta \Delta(\Hb))}{\sinh(2\eta)} \,.
\ee

Not only $\tb(u)$ but also $t(u)$ (\ref{teps1}) has the QG symmetry
\be
\left[ \Delta_{N}(\Hb) \,, t(u) \right] = 0 \,, \qquad
\left[ \Delta_{N}(\E^{\pm}) \,, t(u) \right] = 0 \,, 
\label{QGsym}
\ee
which is consistent with the factorization identity (\ref{tttrltneps1b}).

\vspace{0.4cm}

\noindent
{\bf $Z_{2}$ symmetry}

The open-chain transfer matrix $t(u)$ (\ref{teps1}) has the property
\be
\C\, t(u)\, 
\C
= t(u+ i\pi) \,,
\label{Ct}
\ee
where $\C$ is given by (\ref{Cbb}), similarly to the closed-chain 
transfer matrix (\ref{Ctclosed}). Indeed, the monodromy matrix
identities (\ref{Tdiscrete}) and
\be
\C\, \widehat{T}_{\bar{0}}(u)\, 
\C
= (-1)^N \, \widehat{T}_{\bar{0}}(u+ i\pi) 
\label{Tdiscrete2}
\ee
imply 
\be
\C\, 
T_{\bar{0}}(u)\, \widehat{T}_{\bar{0}}(u+ i\pi)\,
\C
= T_{\bar{0}}(u+ i\pi)\, \widehat{T}_{\bar{0}}(u) \,.
\label{Ctproof}
\ee
Multiplying both sides of (\ref{Ctproof}) by $M_{\bar{0}}$ and 
tracing over the auxiliary space $\bar{0}$, we obtain the desired result (\ref{Ct}).

One consequence of the property (\ref{Ct}) is that the $D^{(2)}_{2}$ 
open-chain transfer matrix $\tb(u)$ has the $Z_{2}$ symmetry
\be
\C\, \tb(u)\,
\C
= \tb(u) \,.
\label{Csymmetry}
\ee
Indeed, we see from the factorization identity (\ref{tttrltneps1b}) 
that
\begin{align}
\C\, \tb(u)\,
\C
&= \phi(u)\, \C\, 
t(u+i\pi)\, t(u)\,
\C \non\\
&= \phi(u)\, t(u)\, t(u+i\pi) = \tb(u)
\,,
\end{align}
where we have passed to the second line using (\ref{Ct}) 
and the $2 i \pi$-periodicity of $t(u)$;
the final equality follows from the commutativity property 
(\ref{commutativity}).
The $Z_{2}$ symmetry of the open-chain transfer matrix
(\ref{Csymmetry}) was first noted in \cite{Robertson:2020imc}.

The QG and $Z_{2}$ generators commute
\be
\left[ \C\,,  
 \Delta_{N}(\Hb) \right] = 0 \,, \qquad
\left[ \C \,, 
\Delta_{N}(\E^{\pm}) \right] = 0 \,.
\ee

\subsubsection{Degeneracies}\label{sec:degopen}

For real values of $\eta$, the degeneracies of the $D^{(2)}_{2}$
open-chain transfer matrix $\tb(u)$ are higher than expected from QG symmetry
alone, as discussed in \cite{Nepomechie:2017hgw, Nepomechie:2018wzp,
Nepomechie:2019tbr}.  These higher degeneracies can now be fully
explained using the above $Z_{2}$ symmetry.

Realizing from (\ref{topen}) and (\ref{Boperator}) that the 
double-row monodromy matrix is given here by
\be
{\cal U}_{0}(u) = T_{0}(u)\, \widehat{T}_{0}(u+i\pi) = \left(\begin{array}{cc}
* & {\cal B}(u) \\
* & * 
\end{array} \right) \,,
\ee
we see from (\ref{Ctproof}) that the $Z_{2}$ symmetry shifts the
argument of the ${\cal B}$-operator by $i\pi$
\be
\C\, 
{\cal B}(u)\,
\C
= {\cal B}(u+ i\pi) \,,
\ee 
similarly to the closed-chain case (\ref{Bshiftclosed}).
The Bethe states (\ref{Bethestateopen}) therefore transform as follows
\be
\C\,
| v_{1} \cdots v_{m} \rangle =  | v_{1}+ i\pi \cdots v_{m}+ i\pi 
\rangle \,.
\ee
In other words, under the $Z_{2}$ symmetry, each of the 
Bethe roots $v_{k}$ (or, equivalently, $u_{k}$) is shifted by 
$i\pi$.

The property (\ref{Ct}) implies that $t(u)$ and $t(u+ i\pi)$ are
related by a unitary transformation (at least for real values of
$\eta$, since $C$ is involutory and symmetric), and therefore have the
same spectrum.  Hence, if $\chi(u)$ is an eigenvalue of $t(u)$, then
$\chi(u+ i\pi)$ is also an eigenvalue of $t(u)$.  Thus, if $Q(u)$
satisfies the TQ-equation, then $Q(u+ i\pi)$ also satisfies the
TQ-equation, as follows simply from performing the shift $u \mapsto u
+ i \pi$ in (\ref{chiopeneps1b}).  Hence, given a set of Bethe roots
$\{ u_{k} \}$, there are only two possibilities for the corresponding
Q-function (\ref{Qeps1}):

\begin{itemize}
	\item $Q(u+ i\pi)= Q(u)$, in which case the corresponding Bethe state is an 
	eigenstate of the $Z_{2}$ symmetry $\C$. The Bethe state is a highest-weight state of a 
	representation of the QG with odd
dimension \cite{Nepomechie:2017hgw, Nepomechie:2018wzp, 
Nepomechie:2019tbr}; hence, the corresponding eigenvalue has odd 
degeneracy.
	
    \item $Q(u+ i\pi)\ne Q(u)$, in which case the Bethe states 
	corresponding to $Q(u)$ and $Q(u+ i\pi)$ are mapped into each 
	other by the $Z_{2}$ symmetry $\C$. It follows from (\ref{Csymmetry}) 
	that the two Bethe states have the same eigenvalue of 
$\tb(u)$, which means that they are degenerate. The degeneracy of the 
corresponding eigenvalue is doubled, and is therefore even.
	
\end{itemize}

\subsubsection{Hamiltonian}

For an open-chain transfer matrix $t(u)$ constructed with a regular 
R-matrix (\ref{reg}) and with all inhomogeneity parameters 
$\{\theta_{l}\}$ 
set to zero (i.e., a homogeneous spin chain), a local Hamiltonian can 
be obtained simply from $t'(0)$ \cite{Sklyanin:1988yz}. However, since the transfer 
matrix (\ref{teps1}) corresponds to a spin chain with inhomogeneities 
at alternate sites, $t'(0)$ is not local. Nevertheless, a local 
Hamiltonian can be obtained from $\frac{d}{du}\log t(u) 
\Big\vert_{u=0} = t^{-1}(0)\, t'(0)$, which is the familiar 
prescription for \emph{periodic}
homogeneous chains. 

For the $A^{(1)}_{1}$ transfer matrix (\ref{teps1}), we obtain
\be
t^{-1}(0)\, t'(0) = \frac{1}{\sinh(2\eta)} {\cal H} + c\, \id \,,
\ee
where, in terms of Temperley-Lieb operators \cite{Temperley:1971iq}
\be
\eb = \left(\begin{array}{cccc}
0 & 0        & 0        & 0\\
0 & e^{\eta} & 1        & 0\\
0 & 1        & e^{-\eta} & 0\\
0 & 0        & 0        & 0
\end{array}\right) \,,
\label{TL}
\ee
the Hamiltonian ${\cal H}$ is given by
\be
{\cal H} = 2 \cosh(\eta) \sum_{j=1}^{2N-1} \eb_{j} - 
\sum_{j=1}^{2N-2}\left( \eb_{j}\, \eb_{j+1} + \eb_{j+1}\, \eb_{j} \right) \,,
\label{Hameps1}
\ee
and $c = -\frac{4}{\sinh(4\eta)}\left(N \cosh^{2}(2\eta) + 
\sinh^{2}(\eta) \right)$.

This Hamiltonian coincides with the Hamiltonian obtained from the 
$D^{(2)}_{2}$ transfer matrix $\tb(u)$ \cite{Robertson:2020imc}.
This fact can be understood from the factorization identity 
(\ref{tttrltneps1b}). We first observe that $\tb'(0) \propto \id$, since
the scalar prefactor $\phi(u)$ vanishes at $u=0$, and also
$t(i \pi)\, t(0) \propto \id$. Indeed, 
\begin{align}
t(0) &= \left(\tfrac{\sinh(2\eta)}{2} \right)^{2N-1} 
\sinh(\eta)\, \cosh(2\eta) \, {\mathcal W} \,, 
\non\\
\qquad 
t(i \pi) &= \left(\tfrac{\sinh(2\eta)}{2} \right)^{2N-1} \sinh(\eta)\, \cosh(2\eta) \,
{\mathcal W}^{-1} \,,
\label{tspecial}
\end{align}
where ${\mathcal W}$ is defined by 
\be
{\mathcal W} = C_{\bar{1} \bar{\bar{1}}}\,  C_{\bar{\bar{1}} \bar{2}}
\ldots C_{\bar{\bar{N-1}}\, \bar{N}}\, C_{\bar{N} \bar{\bar{N}}}\,.
\label{calW}
\ee
Hence, in order to obtain a nontrivial Hamiltonian from $\tb(u)$, one
must differentiate twice, as already noted in \cite{Robertson:2020imc}.
The factorization identity (\ref{tttrltneps1b}) implies
\be
\tb''(0) = 2 \phi'(0) \left[ t'(i \pi)\, t(0) + t(i \pi)\, t'(0) \right]  
+ \text{ const} \,.
\ee
Since $t'(i \pi)\, t(0) = t(i \pi)\, t'(0)$, we conclude that 
$\tb''(0) \propto {\cal H} + \text{ const}$, with ${\cal H}$ given by 
(\ref{Hameps1}).


\subsection{The case $\varepsilon=0$}\label{sec:eps0} 

We now consider the case $\varepsilon=0$, which is similar to the previous case, except for 
one key difference. The $D^{(2)}_{2}$ transfer matrix $\tilde{\tb}(u)$ (\ref{ttopen}) 
again satisfies 
\be
\tilde{\tb}(u) = \B\, 
\tb(u)\,
\B \,,
\label{tttrltneps0a}
\ee
but $\tb(u)$ now satisfies the factorization identity
\be
\tb(u) = \phi(u+\tfrac{i \pi}{2})\, t(u+i\pi)\, t(u) \,,
\label{tttrltneps0b}
\ee
where $t(u)$ is an $A^{(1)}_{1}$ open-chain transfer matrix defined by
\be
t(u) = \tr_{\bar{0}} \Big\{ 
M_{\bar{0}}\, 
T_{\bar{0}}(u)\, 
\widehat{T}_{\bar{0}}(u)\Big\}\,.
\label{teps0}
\ee
As before, $\phi(u)$ is defined in (\ref{tttrltneps1b}),
and $T_{\bar{0}}(u)$ and $\widehat{T}_{\bar{0}}(u)$
are defined in (\ref{monodromyT}) and (\ref{hatmonodromyT}), respectively.
The proof of this factorization identity is also presented in Appendix 
\ref{sec:factopen}.

Note that the transfer matrix (\ref{teps0}), in contrast with the
previous case (\ref{teps1}), does \emph{not} have any shift 
in the argument of $\widehat{T}$ (compared with $T$). Indeed, 
the transfer matrix (\ref{teps0}) is of the standard form 
\cite{Sklyanin:1988yz}. This is the key difference, alluded to above, 
between the $\varepsilon=1$ and $\varepsilon=0$ cases.

\subsubsection{Algebraic Bethe ansatz}\label{sec:ABAeps0} 

We can immediately diagonalize the transfer matrix (\ref{teps0})
using our previous results (\ref{Boperator})-(\ref{chijopen}):  
simply set (as before) $L=2N$ and choose the 
inhomogeneities $\{\theta_{l}\}$ as in (\ref{inhomogparams}), but now set 
the shift $u_{0}= 0$. Hence, the 
Bethe states (\ref{Bethestateopen}) with these 
parameter values are eigenstates of the transfer matrix (\ref{teps0}), with 
corresponding eigenvalues given by
\be
\chi(u) =  
2^{-2N} \left[
\frac{\sinh(u-2\eta)}{\sinh(u-\eta)}\frac{Q(u+\eta)}{Q(u-\eta)}\sinh^{2N}(u-2\eta)
+\frac{\sinh(u)}{\sinh(u-\eta)}\frac{Q(u-3\eta)}{Q(u-\eta)}\sinh^{2N}(u)\right] \,, 
\label{chiopeneps0b}
\ee
with
\be
Q(u) = \prod_{k=1}^{m} \sinh(\tfrac{1}{2}(u-u_{k}))\, 
\sinh(\tfrac{1}{2}(u+u_{k})) \,,
\ee
provided that $u_{j} \equiv v_{j}-\eta$ satisfy the Bethe equations
\be
\left(\frac{\sinh(u_{j}+\eta)}{\sinh(u_{j}-\eta)}\right)^{2N} = 
\prod_{k=1; k\ne j}^{m} 
\frac{\sinh(\tfrac{1}{2}(u_{j}-u_{k})+\eta)\, 
\sinh(\tfrac{1}{2}(u_{j}+u_{k})+\eta)\,}
{\sinh(\tfrac{1}{2}(u_{j}-u_{k})-\eta)\, 
\sinh(\tfrac{1}{2}(u_{j}+u_{k})-\eta)} \,.
\label{BEopeneps0}
\ee

Returning to the $D^{(2)}_{2}$ open-chain transfer matrix 
$\tilde{\tb}(u)$ (\ref{ttopen}) with $\varepsilon=0$, we conclude from the 
factorization identity (\ref{tttrltneps0a})-(\ref{tttrltneps0b}) that its Bethe states are given by 
\be
\B\,  | v_{1} \cdots v_{m} \rangle \,,
\ee
where the vectors $| v_{1} \cdots v_{m} \rangle$ are given by (\ref{Bethestateopen}),
and $\B$ is given by (\ref{Bbb}),
which is a new result. 
Moreover, the corresponding eigenvalues $\Lambda(u)$ are given by
\be
\Lambda(u) = \phi(u+\tfrac{i \pi}{2})\, \chi(u)\, \chi(u+i\pi) \,,
\ee
where $\chi(u)$ is given by (\ref{chiopeneps0b}), and the associated Bethe 
equations are given by (\ref{BEopeneps0}). The Bethe equations agree 
with those obtained by coordinate Bethe ansatz in 
\cite{Martins:2000xie}; the transfer-matrix eigenvalues and Bethe 
equations agree with those obtained 
by analytical Bethe ansatz in \cite{Nepomechie:2017hgw, Nepomechie:2018nvl}.

The symmetries and degeneracies for the $\varepsilon=0$ case are the 
same as for $\varepsilon=1$.

\subsubsection{Hamiltonian}

From the $A^{(1)}_{1}$ transfer matrix (\ref{teps0}), we can generate 
two distinct local Hamiltonians, by evaluating its logarithmic 
derivative at $0$ and at $i\pi$
\begin{align}
t^{-1}(0)\, t'(0) &= \frac{2}{\sinh(2\eta)} {\cal H}^{(1)} + c\, \id 
\,, \non \\
t^{-1}(i\pi)\, t'(i\pi) &= \frac{2}{\sinh(2\eta)} {\cal H}^{(2)} + c\, \id 
\,,
\end{align}
where, in terms of the Temperley-Lieb operators (\ref{TL}), the 
Hamiltonians ${\cal H}^{(1)}$ and ${\cal H}^{(2)}$ are given by
\begin{align}
{\cal H}^{(1)} &= -\frac{1}{\cosh(\eta)}\eb_{1} +  \cosh(\eta) \sum_{j=1}^{2N-1} \eb_{j} - 
\sum_{j=2;\, j= \text{even}}^{2N-2}\left( \eb_{j}\, \eb_{j+1} + \eb_{j+1}\, 
\eb_{j} \right) \,, \non \\
{\cal H}^{(2)} &= -\frac{1}{\cosh(\eta)}\eb_{2N-1} +  \cosh(\eta) \sum_{j=1}^{2N-1} \eb_{j} - 
\sum_{j=1;\, j= \text{odd}}^{2N-3}\left( \eb_{j}\, \eb_{j+1} + \eb_{j+1}\, 
\eb_{j} \right) \,,
\label{Hameps0}
\end{align}
and $c=\frac{1}{\sinh(2\eta)}\left(1 - 2 N \cosh(2\eta) \right)$.

We can use the factorization identity (\ref{tttrltneps0b}) to 
relate these Hamiltonians to the Hamiltonian ${\cal H}$ coming from the 
$D^{(2)}_{2}$ transfer matrix $\tb(u)$. We obtain, up to an additive constant,
\begin{align}
\tb'(0) & \propto  t(i \pi)\, t'(0) + t(0)\, t'(i \pi)  \non\\
        & \propto t^{-1}(0)\, t'(0) + t^{-1}(i \pi)\, t'(i \pi) \,,
\end{align}
since $t(i \pi)\, t(0) \propto \id$. Hence, $\tb'(0) \propto {\cal H} + \text{ const}$, 
with ${\cal H} = {\cal H}^{(1)} + {\cal H}^{(2)}$, in agreement with \cite{Robertson:2020eri}. 

\section{An XXZ-like open spin chain with general $u_{0}$}\label{sec:gen}

The open spin chain with transfer matrix (\ref{topen}) has the exact Bethe ansatz
solution (\ref{Boperator})-(\ref{chijopen}) for any values of $u_{0}$ 
and $\{\theta_{l}\}$. For such generic values, this model does not 
have a local Hamiltonian. However, a local Hamiltonian \emph{can} be 
obtained for general values of $u_{0}$ if we choose the bulk
inhomogeneities to be $-u_{0}$ at alternate sites.
Indeed, let us set
\be
\theta_{l} = \begin{cases}
-u_{0} & \text{for $l=$ odd}\\
0 & \text{for $l=$ even}
\end{cases} \,,
\label{inhomogparams2}
\ee
where $u_{0}$ is arbitrary. We then obtain from (\ref{topen})
\be
t^{-1}(0)\, t'(0) = \frac{1}{\sinh(\eta)} {\cal H} + c(u_{0})\, \id \,,
\ee
where the Hamiltonian ${\cal H}$ is given in terms of Temperley-Lieb 
operators (\ref{TL}) by
\begin{align}
{\cal H} &= \sum_{j=1}^{2N-1} \eb_{j} - 
\frac{1}{2}\sinh(\tfrac{u_{0}}{2})\Bigg\{
\sum_{j=2;\, j= \text{even}}^{2N-2}\left( 
\frac{1}{\sinh(\frac{u_{0}}{2}+\eta)} \eb_{j}\, \eb_{j+1}
+ \frac{1}{\sinh(\frac{u_{0}}{2}-\eta)} \eb_{j+1}\, \eb_{j} \right) \non \\
&\qquad\qquad +
\sum_{j=1;\, j= \text{odd}}^{2N-3}\left( 
\frac{1}{\sinh(\frac{u_{0}}{2}-\eta)} \eb_{j}\, \eb_{j+1}
+ \frac{1}{\sinh(\frac{u_{0}}{2}+\eta)} \eb_{j+1}\, \eb_{j} \right)\Bigg\}\,,
\label{Hamgen}
\end{align}
and the constant $c(u_{0})$ is given by
\be
c(u_{0}) = -\frac{\sinh(\frac{u_{0}}{2}-2\eta)}
{\sinh(\eta)\, \sinh(\frac{u_{0}}{2}-\eta)}N + 
\frac{\sinh(\eta)}
{\sinh(\frac{u_{0}}{2}-\eta)\, \sinh(\frac{u_{0}}{2}-2\eta)} \,.
\ee
(We remark that $t^{-1}(-u_{0})\, t'(-u_{0})$ gives the same Hamiltonian 
(\ref{Hamgen}) with the constant $c(-u_{0})$.) This Hamiltonian becomes 
proportional to (\ref{Hameps1}) for $u_{0}= i \pi$. For $u_{0} 
\rightarrow 0$, the model reduces to a QG-invariant open XXZ chain.

To obtain the above results, it is helpful to introduce a 
generalization of the matrix $C$ (\ref{Cmat}), namely, 
\be
C(u_{0}) = -\frac{1}{\sinh(\frac{u_{0}}{2}-\eta)} {\cal P} 
R(u_{0})\,, \qquad C(u_{0})\, C(-u_{0}) = \id \,,
\label{Cmatgen}
\ee
which reduces to $C$ (\ref{Cmat}) for $u_{0} = \pm i \pi$. Then, similarly to 
(\ref{tspecial}), we find 
\begin{align}
t(0) &= \sinh^{2N} (\eta) \,
\sinh^{2N-1}\left(\frac{u_0}{2}-\eta\right)\, \sinh\left(\frac{u_0}{2}-2\eta\right) \, 
{\mathcal W}(u_{0}) \,, 
\non\\
\qquad 
t(-u_0) &= \sinh^{2N} (\eta) \,
\sinh^{2N-1}\left(\frac{u_0}{2}+\eta\right)\, \sinh\left(\frac{u_0}{2}+2\eta\right) \,
{\mathcal W}^{-1}(u_{0}) \,,
\end{align}
where
\be
{\mathcal W}(u_{0}) =
C_{\bar{1} \bar{\bar{1}}}(u_{0})\, C_{\bar{\bar{1}} \bar{2}}(u_{0})
\ldots C_{\bar{\bar{N-1}}\, \bar{N}}(u_{0})\, C_{\bar{N} 
\bar{\bar{N}}}(u_{0}) \,.
\ee

For the choice (\ref{inhomogparams2}) of inhomogeneities,
the Bethe states (\ref{Bethestateopen}) are eigenstates of the transfer matrix 
(\ref{topen}), with 
corresponding eigenvalues given by
\begin{align}
\chi(u) &=  
\frac{\sinh(u+\frac{u_{0}}{2}-2\eta)}{\sinh(u+\frac{u_{0}}{2}-\eta)}
\frac{q(u+2\eta)}{q(u)}
\left[\sinh(\tfrac{1}{2}(u+u_{0})-\eta)\,
\sinh(\tfrac{u}{2}-\eta)\right]^{2N} \non\\
&\qquad
+\frac{\sinh(u+\frac{u_{0}}{2})}{\sinh(u+\frac{u_{0}}{2}-\eta)}
\frac{q(u-2\eta)}{q(u)}
\left[\sinh(\tfrac{1}{2}(u+u_{0}))\,
\sinh(\tfrac{u}{2})\right]^{2N} \,, 
\label{chiopengen}
\end{align}
with $q(u)$ given by (\ref{qfunc}),
provided that $\{ v_{j} \}$ satisfy the Bethe equations
\begin{align}
&\left[
\frac{\sinh(\tfrac{1}{2}(v_{j}+u_{0}))\,\sinh(\tfrac{v_{j}}{2})}
{\sinh(\tfrac{1}{2}(v_{j}+u_{0})-\eta)\, \sinh(\tfrac{v_{j}}{2}-\eta)}
\right]^{2N} \non\\
&\qquad = 
\prod_{k=1; k\ne j}^{m}
\frac{\sinh(\tfrac{1}{2}(v_{j}-v_{k})+\eta)\, 
\sinh(\tfrac{1}{2}(v_{j}+v_{k}+u_{0}))}
{\sinh(\tfrac{1}{2}(v_{j}-v_{k})-\eta)\, 
\sinh(\tfrac{1}{2}(v_{j}+v_{k}+u_{0})-2\eta)} \,.
\label{BAEopengen}
\end{align}
In terms of $u_{j} \equiv v_{j}-\eta$, these Bethe equations take a more symmetric form
\begin{align}
&\left[
\frac{\sinh(\tfrac{1}{2}(u_{j}+u_{0})+\tfrac{\eta}{2})\,
\sinh(\tfrac{u_{j}}{2}+\tfrac{\eta}{2})}
{\sinh(\tfrac{1}{2}(u_{j}+u_{0})-\tfrac{\eta}{2})\, 
\sinh(\tfrac{u_{j}}{2}-\tfrac{\eta}{2})}
\right]^{2N} \non\\
&\qquad = 
\prod_{k=1; k\ne j}^{m}
\frac{\sinh(\tfrac{1}{2}(u_{j}-u_{k})+\eta)\, 
\sinh(\tfrac{1}{2}(u_{j}+u_{k}+u_{0})+\eta)}
{\sinh(\tfrac{1}{2}(u_{j}-u_{k})-\eta)\, 
\sinh(\tfrac{1}{2}(u_{j}+u_{k}+u_{0})-\eta)} \,.
\end{align}
For $u_{0}=i\pi$, these equations reduce to (\ref{BEopeneps1}).
Alternatively, in terms of $u_{j} \equiv v_{j}-\eta+\frac{u_{0}}{2}$, 
the Bethe equations (\ref{BAEopengen}) take the form
\begin{align}
&\left[
\frac{\sinh(\tfrac{1}{2}(u_{j}+\frac{u_{0}}{2})+\tfrac{\eta}{2})\,
\sinh(\tfrac{1}{2}(u_{j}-\frac{u_{0}}{2})+\tfrac{\eta}{2})}
{\sinh(\tfrac{1}{2}(u_{j}+\frac{u_{0}}{2})-\tfrac{\eta}{2})\, 
\sinh(\tfrac{1}{2}(u_{j}-\frac{u_{0}}{2})-\tfrac{\eta}{2})}
\right]^{2N} \non\\
&\qquad = 
\prod_{k=1; k\ne j}^{m}
\frac{\sinh(\tfrac{1}{2}(u_{j}-u_{k})+\eta)\, 
\sinh(\tfrac{1}{2}(u_{j}+u_{k})+\eta)}
{\sinh(\tfrac{1}{2}(u_{j}-u_{k})-\eta)\, 
\sinh(\tfrac{1}{2}(u_{j}+u_{k})-\eta)} \,.
\end{align}
We note that these Bethe equations are an ``open-chain version'' of the closed-chain Bethe
equations (3.4) in \cite{Frahm:2013cma}. We also note that the 
transfer matrix has the QG symmetry (\ref{QGgens})-(\ref{QGsym}) for 
any value of $u_{0}$.

We have considered here an integrable model based on the transfer matrix 
(\ref{topen}) with an arbitrary value of $u_{0}$. It should be possible to 
generalize this model by introducing general K-matrices, 
as noted in footnote \ref{foot}. However, this will generally result 
in the breaking of QG symmetry. 

\section{Discussion}\label{sec:discuss}

We have exploited the factorization of the $D^{(2)}_{2}$ R-matrix 
into a product of $A^{(1)}_{1}$ R-matrices 
(\ref{D22Ra})-(\ref{D22Rb}) to derive corresponding factorization 
identities for the transfer matrices of both closed and open spin chains, 
see (\ref{tttrltna})-(\ref{tttrltnb}), 
(\ref{tttrltneps1a})-(\ref{tttrltneps1b}) and 
(\ref{tttrltneps0a})-(\ref{tttrltneps0b}). We have used these factorization 
identities to solve the models by algebraic Bethe ansatz. In 
particular, we have constructed the Bethe states of these models, 
which heretofore had not been known. 
These constructions should be useful for computing scalar products and correlation 
functions. Moreover, we have proved previously-proposed expressions for the 
models' eigenvalues and Bethe equations. The 
interesting degeneracies exhibited by the QG-invariant open chains 
for real values of $\eta$ have now also been 
explained.

In the course of this work, we have uncovered a new integrable
XXZ-like open spin chain, with transfer matrix (\ref{topen}), which
depends on a continuous parameter $u_{0}$.  
We have interpreted this parameter as the rapidity of the boundary.
For inhomogeneities $-i\pi$ at alternate sites (\ref{inhomogparams}),
this model continuously
interpolates between the cases $\varepsilon=0$ ($u_{0}=0$) and
$\varepsilon=1$ ($u_{0}=i \pi$).  
For inhomogeneities $-u_{0}$ at alternate sites (\ref{inhomogparams2}),
this model has a local Hamiltonian (\ref{Hamgen}) for general values of $u_{0}$.
We conjecture that, for the parameters $\eta$ and $u_{0}$ in suitable 
domains, the continuum limit of the latter model is a \emph{non-compact}
boundary conformal field theory, as is the case for $u_{0}=i \pi$ \cite{Robertson:2020imc}, 
see also \cite{Jacobsen:2005xz, Ikhlef:2008zz, Ikhlef:2011ay,
Candu:2013fva, Frahm:2013cma, Bazhanov:2019xvy}.

\section*{Acknowledgments}
We thank Nicolas Cramp\'e, Tamas Gombor, Rodrigo Pimenta and
especially Niall Robertson for valuable correspondence and/or 
discussions. We benefitted greatly from access to the latter's
unpublished thesis, part of which is included in \cite{Robertson:2020imc}. 
A.L.R. is supported by Grant 404 No. 18/EPSRC/3590. 

\appendix

\section{Factorization identities for open chains}\label{sec:factopen}

We present here the derivations of the open-chain factorization identities 
(\ref{tttrltneps1a})-(\ref{tttrltneps1b}) and 
(\ref{tttrltneps0a})-(\ref{tttrltneps0b}).
The initial steps of the derivations are the same for both cases.  We
then focus on the case $\varepsilon=1$ in Sec.  \ref{sec:facteps1},
followed by the case $\varepsilon=0$ in Sec.  \ref{sec:facteps0}.

We begin the derivation of the factorization identities by 
substituting into the formula for the open-chain transfer matrix (\ref{ttopen}) 
the factorized expressions for the monodromy matrices, namely, 
(\ref{TTTrltna})-(\ref{TTTrltnb}) for $\tilde{\T}_{\bar{0} 
\bar{\bar{0}}}(u)$, and (\ref{hatTTTrltna})-(\ref{hatTTTrltnb}) for 
$\widehat{\tilde{\T}}_{\bar{0} \bar{\bar{0}}}(u)$. In this way, we 
obtain
\be
\tilde{\tb}(u) = \B\, 
\tb(u)\,
\B\,,
\ee
where
\begin{align}
\tb(u) &= 2^{8N}\, 
\tr_{\bar{0} \bar{\bar{0}}} \Bigg\{
\tilde{\K}^{L}_{\bar{0} \bar{\bar{0}}}(u)\, 
B_{\bar{0} \bar{\bar{0}}}\, 
\Big[ \C\, 
T_{\bar{0}}(u)\, 
T_{\bar{\bar{0}}}(u-i\pi)\, 
\C \Big] \non\\
&\qquad\times
B_{\bar{0} \bar{\bar{0}}}\,
\tilde{\K}^{R}_{\bar{0} \bar{\bar{0}}}(u)\, 
B_{\bar{0} \bar{\bar{0}}}\,
C_{\bar{0} \bar{\bar{0}}}\, 
\widehat{T}_{\bar{\bar{0}}}(u+i\pi)\, 
\widehat{T}_{\bar{0}}(u)\, 
C_{\bar{0} \bar{\bar{0}}}\,
B_{\bar{0} \bar{\bar{0}}}\Bigg\} \,.
\label{step1}
\end{align}
Using the first identity in ({\ref{Tdiscrete}),
we see that the product of terms within square brackets in 
(\ref{step1}) is equal to 
$T_{\bar{0}}(u + i\pi)\, T_{\bar{\bar{0}}}(u)\,. $
The expression for $\tb(u)$ in 
(\ref{step1}) therefore reduces to
\begin{align}
\tb(u) &= 2^{8N}\, 
\tr_{\bar{0} \bar{\bar{0}}} \Bigg\{
B_{\bar{0} \bar{\bar{0}}}\,
\tilde{\K}^{L}_{\bar{0} \bar{\bar{0}}}(u)\, 
B_{\bar{0} \bar{\bar{0}}}\, 
T_{\bar{0}}(u + i\pi)\, 
T_{\bar{\bar{0}}}(u)\, \non\\
&\qquad \times
B_{\bar{0} \bar{\bar{0}}}\,
\tilde{\K}^{R}_{\bar{0} \bar{\bar{0}}}(u)\, 
B_{\bar{0} \bar{\bar{0}}}\,
C_{\bar{0} \bar{\bar{0}}}\, 
\widehat{T}_{\bar{\bar{0}}}(u+i\pi)\, 
\widehat{T}_{\bar{0}}(u)\, 
C_{\bar{0} \bar{\bar{0}}}\Bigg\} \,.
\label{step2}
\end{align}

\subsection{The case $\varepsilon=1$}\label{sec:facteps1}

We now focus on the case $\varepsilon=1$. The key step, having 
already expressed the $\tilde{\R}$'s in terms of $R$'s, is to also express 
the $\tilde{\K}$'s in terms of $R$'s. Remarkably,
the right K-matrix (\ref{KR}) with $\varepsilon=1$ satisfies the identity
\be
B_{\bar{0} \bar{\bar{0}}}\,
\tilde{\K}^{R}_{\bar{0} \bar{\bar{0}}}(u)\, 
B_{\bar{0} \bar{\bar{0}}} = \frac{1}{\sinh(u+\eta)} 
{\cal P}_{\bar{0} \bar{\bar{0}}}\, 
R_{\bar{0} \bar{\bar{0}}}(2u) \,.
\label{KReps1}
\ee
Eq. (\ref{step2}) therefore further simplifies to
\begin{align}
\tb(u) &= \frac{2^{8N}}{\sinh(u+\eta)} \, 
\tr_{\bar{0} \bar{\bar{0}}} \Bigg\{
B_{\bar{0} \bar{\bar{0}}}\,
\tilde{\K}^{L}_{\bar{0} \bar{\bar{0}}}(u)\, 
B_{\bar{0} \bar{\bar{0}}}\, 
T_{\bar{0}}(u + i\pi)\, 
T_{\bar{\bar{0}}}(u)\, \non\\
&\qquad \times
{\cal P}_{\bar{0} \bar{\bar{0}}}\, 
R_{\bar{0} \bar{\bar{0}}}(2u)\,
C_{\bar{0} \bar{\bar{0}}}\, 
\widehat{T}_{\bar{\bar{0}}}(u+i\pi)\, 
\widehat{T}_{\bar{0}}(u)\, 
C_{\bar{0} \bar{\bar{0}}}\Bigg\} \,.
\label{step3a}
\end{align}
The product of terms on the second line of (\ref{step3a}) can be 
simplified as follows:
\begin{align}
& \Big[ {\cal P}_{\bar{0} \bar{\bar{0}}}\, 
R_{\bar{0} \bar{\bar{0}}}(2u)\Big]\,
C_{\bar{0} \bar{\bar{0}}}\, 
\widehat{T}_{\bar{\bar{0}}}(u+i\pi)\, 
\widehat{T}_{\bar{0}}(u)\, 
C_{\bar{0} \bar{\bar{0}}} \non\\
&\qquad =
R_{\bar{\bar{0}} \bar{0}}(2u)\,
\Big[{\cal P}_{\bar{0} \bar{\bar{0}}}\, 
C_{\bar{0} \bar{\bar{0}}}\, 
\widehat{T}_{\bar{\bar{0}}}(u+i\pi)\, 
\widehat{T}_{\bar{0}}(u)\Big]\, 
C_{\bar{0} \bar{\bar{0}}} \non\\
&\qquad
= R_{\bar{\bar{0}} \bar{0}}(2u)\,
\widehat{T}_{\bar{0}}(u)\,
\widehat{T}_{\bar{\bar{0}}}(u+i\pi)\,
{\cal P}_{\bar{0} \bar{\bar{0}}}\, 
\Big[ C_{\bar{0} \bar{\bar{0}}}\,
C_{\bar{0} \bar{\bar{0}}} \Big] \non\\
&\qquad=
R_{\bar{\bar{0}} \bar{0}}(2u)\,
\widehat{T}_{\bar{0}}(u)\,
\widehat{T}_{\bar{\bar{0}}}(u+i\pi)\,
{\cal P}_{\bar{0} \bar{\bar{0}}} \,,
\label{note}
\end{align}
where square brackets are used to indicate the terms to be 
transformed in the subsequent step. In passing to the third line of (\ref{note}), we have used the identity
\be
{\cal P}_{\bar{0} \bar{\bar{0}}}\, 
C_{\bar{0} \bar{\bar{0}}}\, 
\widehat{T}_{\bar{\bar{0}}}(u+i\pi)\, 
\widehat{T}_{\bar{0}}(u) =
\widehat{T}_{\bar{0}}(u)\,
\widehat{T}_{\bar{\bar{0}}}(u+i\pi)\,
{\cal P}_{\bar{0} \bar{\bar{0}}}\, 
C_{\bar{0} \bar{\bar{0}}} \,,
\ee 
which follows from the fact ${\cal P}\, C \propto R(i \pi)$ (see 
(\ref{Cmat})) and the second relation in (\ref{fundamrltns}).
Eq. (\ref{step3a}) therefore becomes
\begin{align}
\tb(u) &= \frac{2^{8N}}{\sinh(u+\eta)} \, 
\tr_{\bar{0} \bar{\bar{0}}} \Bigg\{
B_{\bar{0} \bar{\bar{0}}}\,
\tilde{\K}^{L}_{\bar{0} \bar{\bar{0}}}(u)\, 
B_{\bar{0} \bar{\bar{0}}}\, \non\\
&\qquad \times
T_{\bar{0}}(u + i\pi)\, 
\Big[ T_{\bar{\bar{0}}}(u)\, 
R_{\bar{\bar{0}} \bar{0}}(2u)\,
\widehat{T}_{\bar{0}}(u) \Big]\,
\widehat{T}_{\bar{\bar{0}}}(u+i\pi)\,
{\cal P}_{\bar{0} \bar{\bar{0}}}
\Bigg\} \,.
\label{step4a}
\end{align}
Using the third relation in (\ref{fundamrltns}),
we arrive at 
\begin{align}
\tb(u) &= \frac{2^{8N}}{\sinh(u+\eta)} \, 
\tr_{\bar{0} \bar{\bar{0}}} \Bigg\{
{\cal P}_{\bar{0} \bar{\bar{0}}}\,
B_{\bar{0} \bar{\bar{0}}}\,
\tilde{\K}^{L}_{\bar{0} \bar{\bar{0}}}(u)\, 
B_{\bar{0} \bar{\bar{0}}}\, \non\\
&\qquad \times
T_{\bar{0}}(u + i\pi)\, 
\widehat{T}_{\bar{0}}(u)\,
R_{\bar{\bar{0}} \bar{0}}(2u)\,
T_{\bar{\bar{0}}}(u)\,
\widehat{T}_{\bar{\bar{0}}}(u+i\pi)\,
\Bigg\} \,.
\label{step5a}
\end{align}
The left K-matrix (\ref{KL}) satisfies, as a consequence of the 
identity for the right K-matrix (\ref{KReps1}), the following 
corresponding identity
\be
{\cal P}_{\bar{0} \bar{\bar{0}}}\,
B_{\bar{0} \bar{\bar{0}}}\,
\tilde{\K}^{L}_{\bar{0} \bar{\bar{0}}}(u)\, 
B_{\bar{0} \bar{\bar{0}}} = -\frac{1}{\sinh(u-3\eta)} 
M_{\bar{0}}\, M_{\bar{\bar{0}}}\, 
R_{\bar{0} \bar{\bar{0}}}(-2u+4\eta) \,.
\ee
Hence, (\ref{step5a}) becomes
\begin{align}
\tb(u) &= -\frac{2^{8N}}{\sinh(u+\eta)\sinh(u-3\eta)} \, 
\tr_{\bar{0} \bar{\bar{0}}} \Bigg\{
M_{\bar{0}}\, 
R_{\bar{0} \bar{\bar{0}}}(-2u+4\eta)\,
M^{-1}_{\bar{0}}\, \non\\
&\qquad \times
\Big[ M_{\bar{0}}\, 
T_{\bar{0}}(u + i\pi)\, 
\widehat{T}_{\bar{0}}(u)\Big]\,
R_{\bar{\bar{0}} \bar{0}}(2u)\,
\Big[T_{\bar{\bar{0}}}(u)\,
\widehat{T}_{\bar{\bar{0}}}(u+i\pi)\,
M_{\bar{\bar{0}}}\Big]\, 
\Bigg\} \,.
\label{step6a}
\end{align}
We next make use of the identity
\begin{align}
\tr_{\bar{0} \bar{\bar{0}}} \Big\{
& M_{\bar{0}}\, 
R_{\bar{0} \bar{\bar{0}}}(-2u+4\eta)\,
M^{-1}_{\bar{0}}\,
F_{\bar{0} a}\,
R_{\bar{\bar{0}} \bar{0}}(2u)\,
G_{\bar{\bar{0}} a}\Big\} \non\\
&\qquad 
= -\sinh u\, \sinh(u-2\eta)\, \tr_{\bar{0} \bar{\bar{0}}} \Big\{ 
F_{\bar{0} a}\,
G_{\bar{\bar{0}} a} \Big\} \,,
\label{FGid}
\end{align}
where $F$ and $G$ are arbitrary, whose proof is as follows:
\begin{align}
&-\sinh u\, \sinh(u-2\eta)\, 
\tr_{\bar{0} \bar{\bar{0}}} \Big\{ 
F_{\bar{0} a}\,
G_{\bar{\bar{0}} a} \Big\} \non\\
&\qquad = 
-\sinh u\, \sinh(u-2\eta)\,
\tr_{\bar{0} \bar{\bar{0}}} \Big\{ 
F_{\bar{0} a}\,
G^{t_{\bar{\bar{0}}}}_{\bar{\bar{0}} a} \Big\} \non \\
&\qquad =
\tr_{\bar{0} \bar{\bar{0}}} \Big\{ 
F_{\bar{0} a}\,
G^{t_{\bar{\bar{0}}}}_{\bar{\bar{0}} a}\, 
R_{\bar{\bar{0}} \bar{0}}^{t_{\bar{\bar{0}}}}(2u)\, M_{\bar{0}}\, 
R_{\bar{0} \bar{\bar{0}}}^{t_{\bar{\bar{0}}}}(-2u+4\eta)\, M^{-1}_{\bar{0}}
\Big\} \non \\
&\qquad =
\tr_{\bar{0} \bar{\bar{0}}} \Big\{ 
M_{\bar{0}}\, 
R_{\bar{0} \bar{\bar{0}}}^{t_{\bar{\bar{0}}}}(-2u+4\eta)\, 
M^{-1}_{\bar{0}}\,
F_{\bar{0} a}\,
G^{t_{\bar{\bar{0}}}}_{\bar{\bar{0}} a}\,
R_{\bar{\bar{0}} \bar{0}}^{t_{\bar{\bar{0}}}}(2u)
\Big\} \non \\
&\qquad =
\tr_{\bar{0} \bar{\bar{0}}} \Big\{ 
M_{\bar{0}}\, 
R_{\bar{0} \bar{\bar{0}}}^{t_{\bar{\bar{0}}}}(-2u+4\eta)\, 
M^{-1}_{\bar{0}}\,
F_{\bar{0} a}\,
\Big[R_{\bar{\bar{0}} \bar{0}}(2u)\,
G_{\bar{\bar{0}} a} \Big]^{t_{\bar{\bar{0}}}}
\Big\} \non \\
&\qquad =
\tr_{\bar{0} \bar{\bar{0}}} \Big\{ 
\Big[M_{\bar{0}}\, 
R_{\bar{0} \bar{\bar{0}}}^{t_{\bar{\bar{0}}}}(-2u+4\eta)\, 
M^{-1}_{\bar{0}}\,
F_{\bar{0} a}\Big]^{t_{\bar{\bar{0}}}}
R_{\bar{\bar{0}} \bar{0}}(2u)\,
G_{\bar{\bar{0}} a}
\Big\} \non \\
&\qquad =
\tr_{\bar{0} \bar{\bar{0}}} \Big\{
M_{\bar{0}}\, 
R_{\bar{0} \bar{\bar{0}}}(-2u+4\eta)\,
M^{-1}_{\bar{0}}\,
F_{\bar{0} a}\,
R_{\bar{\bar{0}} \bar{0}}(2u)\,
G_{\bar{\bar{0}} a}\Big\} \,.
\end{align}
In passing to the third line, we have used the crossing-unitarity 
(\ref{crossunit}) and PT-symmetry (\ref{PTsym}) of $R(u)$. In the 
subsequent step, we have repeatedly used the cyclic property of the trace.

Making use of the identity (\ref{FGid}) in (\ref{step6a}), we finally obtain
\begin{align}
\tb(u) &= \phi(u)\,
\tr_{\bar{0} \bar{\bar{0}}} \Bigg\{ 
\Big[ M_{\bar{0}}\, 
T_{\bar{0}}(u + i\pi)\, 
\widehat{T}_{\bar{0}}(u)\Big]\,
\Big[T_{\bar{\bar{0}}}(u)\,
\widehat{T}_{\bar{\bar{0}}}(u+i\pi)\,
M_{\bar{\bar{0}}}\Big]\, 
\Bigg\} \non\\
& = \phi(u)\,
\tr_{\bar{0}} \Big\{ 
M_{\bar{0}}\, 
T_{\bar{0}}(u + i\pi)\, 
\widehat{T}_{\bar{0}}(u)\Big\}\,  
\tr_{\bar{\bar{0}}} \Big\{ 
T_{\bar{\bar{0}}}(u)\,
\widehat{T}_{\bar{\bar{0}}}(u+i\pi)\,
M_{\bar{\bar{0}}}
\Big\} \non\\
& = \phi(u)\, t(u+i\pi)\, t(u) \,,
\label{step7a}
\end{align}
where $\phi(u)$ and
$t(u)$ are defined in (\ref{tttrltneps1b}) and (\ref{teps1}), 
respectively. This concludes the proof of 
the factorization identity (\ref{tttrltneps1b}).

\subsection{The case $\varepsilon=0$}\label{sec:facteps0}

Let us consider now the case $\varepsilon=0$. Again, the key step is to 
express the $\tilde{\K}$'s in terms of $R$'s. For the right K-matrix (\ref{KR}),
we find
\be
B_{\bar{0} \bar{\bar{0}}}\,
\tilde{\K}^{R}_{\bar{0} \bar{\bar{0}}}(u)\, 
B_{\bar{0} \bar{\bar{0}}} = \frac{i}{\cosh(u+\eta)} 
{\cal P}_{\bar{0} \bar{\bar{0}}}\, 
R_{\bar{0} \bar{\bar{0}}}(2u+i\pi)\,
C_{\bar{0} \bar{\bar{0}}} \,.
\label{KReps0}
\ee
The left K-matrix (\ref{KL}) in turn satisfies
\be
B_{\bar{0} \bar{\bar{0}}}\,
\tilde{\K}^{L}_{\bar{0} \bar{\bar{0}}}(u)\, 
B_{\bar{0} \bar{\bar{0}}} = -\frac{i}{\cosh(u-3\eta)}\,
C_{\bar{0} \bar{\bar{0}}}\,
{\cal P}_{\bar{0} \bar{\bar{0}}}\,
R_{\bar{0} \bar{\bar{0}}}(-2u+4\eta-i\pi)\, 
M_{\bar{0}}\, M_{\bar{\bar{0}}} \,.
\ee
Substituting these results into (\ref{step2}), we obtain
\begin{align}
\tb(u) &= \frac{2^{8N}}{\cosh(u+\eta)\cosh(u-3\eta)} \, 
\tr_{\bar{0} \bar{\bar{0}}} \Bigg\{
{\cal P}_{\bar{0} \bar{\bar{0}}}\,
R_{\bar{0} \bar{\bar{0}}}(-2u+4\eta-i\pi)\, 
M_{\bar{0}}\, M_{\bar{\bar{0}}}\, \non\\
&\qquad \times 
T_{\bar{0}}(u + i\pi)\, 
T_{\bar{\bar{0}}}(u)\, {\cal P}_{\bar{0} \bar{\bar{0}}}\,  
R_{\bar{0} \bar{\bar{0}}}(2u+i\pi)\,
\widehat{T}_{\bar{\bar{0}}}(u+i\pi)\, 
\widehat{T}_{\bar{0}}(u)\, 
\Bigg\} \,.
\label{step3b}
\end{align}
The product of terms on the second line of (\ref{step3b}) can be 
simplified as follows:
\begin{align}
& \Big[ T_{\bar{0}}(u + i\pi)\, 
T_{\bar{\bar{0}}}(u)\, {\cal P}_{\bar{0} \bar{\bar{0}}}\Big] \,  
R_{\bar{0} \bar{\bar{0}}}(2u+i\pi)\,
\widehat{T}_{\bar{\bar{0}}}(u+i\pi)\, 
\widehat{T}_{\bar{0}}(u)\, \non\\
&\qquad =
{\cal P}_{\bar{0} \bar{\bar{0}}}\,
T_{\bar{\bar{0}}}(u + i\pi)\, 
\Big[ T_{\bar{0}}(u)\, 
R_{\bar{0} \bar{\bar{0}}}(2u+i\pi)\,
\widehat{T}_{\bar{\bar{0}}}(u+i\pi)\Big]\, 
\widehat{T}_{\bar{0}}(u)\, \non\\
&\qquad =
{\cal P}_{\bar{0} \bar{\bar{0}}}\,
T_{\bar{\bar{0}}}(u + i\pi)\,
\widehat{T}_{\bar{\bar{0}}}(u+i\pi)\,
R_{\bar{0} \bar{\bar{0}}}(2u+i\pi)\,
T_{\bar{0}}(u)\, 
\widehat{T}_{\bar{0}}(u)\,.
\label{note2}
\end{align}
In passing to the third line of (\ref{note2}), we have used the third 
relation in (\ref{fundamrltns}).
Eq. (\ref{step3b}) therefore becomes
\begin{align}
\tb(u) &= \frac{2^{8N}}{\cosh(u+\eta)\cosh(u-3\eta)} \, 
\tr_{\bar{0} \bar{\bar{0}}} \Bigg\{
\Big[{\cal P}_{\bar{0} \bar{\bar{0}}}\,
R_{\bar{0} \bar{\bar{0}}}(-2u+4\eta-i\pi)\, 
M_{\bar{0}}\, M_{\bar{\bar{0}}}\, {\cal P}_{\bar{0} 
\bar{\bar{0}}}\Big]\, \non\\
&\qquad \times 
T_{\bar{\bar{0}}}(u + i\pi)\,
\widehat{T}_{\bar{\bar{0}}}(u+i\pi)\,
R_{\bar{0} \bar{\bar{0}}}(2u+i\pi)\,
T_{\bar{0}}(u)\, 
\widehat{T}_{\bar{0}}(u)
\Bigg\} \non\\
&= \frac{2^{8N}}{\cosh(u+\eta)\cosh(u-3\eta)} \, 
\tr_{\bar{0} \bar{\bar{0}}} \Bigg\{
R_{\bar{\bar{0}} \bar{0} }(-2u+4\eta-i\pi)\, 
M_{\bar{0}}\, M_{\bar{\bar{0}}}\,  \non\\
&\qquad \times 
T_{\bar{\bar{0}}}(u + i\pi)\,
\widehat{T}_{\bar{\bar{0}}}(u+i\pi)\,
R_{\bar{0} \bar{\bar{0}}}(2u+i\pi)\,
T_{\bar{0}}(u)\, 
\widehat{T}_{\bar{0}}(u)
\Bigg\} \non\\
&= \frac{2^{8N}}{\cosh(u+\eta)\cosh(u-3\eta)} \, 
\tr_{\bar{0} \bar{\bar{0}}} \Bigg\{
M_{\bar{\bar{0}}}\,
R_{\bar{\bar{0}} \bar{0} }(-2u+4\eta-i\pi)\, 
M^{-1}_{\bar{\bar{0}}}\, \non\\
&\qquad \times 
\Big[ M_{\bar{\bar{0}}}\,
T_{\bar{\bar{0}}}(u + i\pi)\,
\widehat{T}_{\bar{\bar{0}}}(u+i\pi)\Big]\,
R_{\bar{0} \bar{\bar{0}}}(2u+i\pi)\,
\Big[ T_{\bar{0}}(u)\, 
\widehat{T}_{\bar{0}}(u)\,
M_{\bar{0}} \Big]\,  
\Bigg\} \,.
\label{step4b}
\end{align}
In passing to the last line, we have used the fact 
$\left[R_{12}(u)\,, M_{1}\, M_{2} \right] = 0$.

Making use of the identity (\ref{FGid}) in (\ref{step4b}), we finally obtain
\begin{align}
\tb(u) &= \phi(u+\tfrac{i \pi}{2})\,
\tr_{\bar{0} \bar{\bar{0}}} \Bigg\{ 
\Big[ M_{\bar{\bar{0}}}\, 
T_{\bar{\bar{0}}}(u + i\pi)\, 
\widehat{T}_{\bar{\bar{0}}}(u + i\pi)\Big]\,
\Big[T_{\bar{0}}(u)\,
\widehat{T}_{\bar{0}}(u)\,
M_{\bar{0}}\Big]\, 
\Bigg\} \non\\
& = \phi(u+\tfrac{i \pi}{2})\,
\tr_{\bar{\bar{0}}} \Big\{ 
M_{\bar{\bar{0}}}\, 
T_{\bar{\bar{0}}}(u+i\pi)\, 
\widehat{T}_{\bar{\bar{0}}}(u+i\pi)\Big\}\,  
\tr_{\bar{0}} \Big\{ 
T_{\bar{0}}(u)\,
\widehat{T}_{\bar{0}}(u)\,
M_{\bar{0}}
\Big\} \non\\
& = \phi(u+\tfrac{i \pi}{2})\, t(u+i\pi)\, t(u) \,,
\label{step7b}
\end{align}
where $\phi(u)$ and
$t(u)$ are defined in (\ref{tttrltneps1b}) and (\ref{teps0}), 
respectively. This concludes the proof of 
the factorization identity (\ref{tttrltneps0b}).


\providecommand{\href}[2]{#2}\begingroup\raggedright\endgroup

\end{document}